\documentclass{article}
\usepackage{graphicx}
\usepackage{amssymb}
\usepackage{amsmath}
\usepackage{amsthm}
\usepackage{epsfig}

\newtheorem{proposition}{Proposition}
\newtheorem{theorem}{Theorem}
\newtheorem{corollary}{Corollary}
\newtheorem{lemma}{Lemma}
\newtheorem{problem}{Problem}

\title{Almost optimal searching of maximal subrepetitions in a word}

\author{
        Roman Kolpakov\\
        Lomonosov Moscow State University,\\  
				Dorodnicyn Computing Centre FRC CSC RAS,\\
				Moscow, Russia\\
        Email: foroman@mail.ru
       }
\date{}

\begin{document}

\maketitle

\begin{abstract}
For $0<\delta <1$ a $\delta$-subrepetition in a word is a factor which exponent 
is less than~2 but is not less than $1+\delta$ (the exponent of the factor is the 
ratio of the factor length to its minimal period). The $\delta$-subrepetition is 
maximal if it cannot be extended to the left or to the right by at least one letter 
with preserving its minimal period. In the paper we propose an algorithm for
searching all maximal $\delta$-subrepetitions in a word of length~$n$ in
$O(\frac{n}{\delta}\log\frac{1}{\delta})$ time (the lower bound for this
time is $\Omega (\frac{n}{\delta})$).
\end{abstract}

Let $w=w[1]w[2]\ldots w[n]$ be an arbitrary word of length $|w|=n$. 
A fragment $w[i]\cdots w[j]$ of~$w$, where $1\le i\le j\le n$, is called 
a {\it factor} of~$w$ and is denoted by $w[i..j]$. Note that this factor 
can be considered either as a word itself or as the fragment $w[i]\ldots w[j]$ 
of~$w$. So for factors we have two different notions of equality: factors 
can be equal as the same fragment of the word~$w$ or as the same word. 
To avoid this ambiguity, we use two different notations: if two factors 
$u$ and $v$ of~$w$ are the same word (the same fragment of~$w$) we will 
write $u=v$ ($u\equiv v$). For any $i=1,\ldots,n$ the factor $w[1..i]$ 
($w[i..n]$) is called a {\it prefix} (a {\it suffix}) of~$w$. By positions 
in~$w$ we mean the order numbers $1, 2,\ldots ,n$ of letters of the word~$w$. 
For any factor~$v\equiv w[i..j]$ of~$w$ the positions $i$ and $j$ are called 
{\it start position} of~$v$ and {\it endind position} of~$v$ and denoted by 
${\rm beg} (v)$ and ${\rm end} (v)$ respectively. For any two factors $u$, $v$ 
of~$w$ the factor $u$ {\it is contained} in~$v$ if ${\rm beg} (v)\le {\rm beg} (u)$ 
and ${\rm end} (u)\le {\rm end} (v)$. Two factors $u$ and $v$ of~$w$
such that ${\rm beg} (u)\le {\rm beg} (v)$  are called {\it overlapped} if
${\rm beg} (v)\le {\rm end} (u)+1$. The {\it intersection} of the overlapped
factors $u$ and $v$ is the factor $w[{\rm beg} (v) .. {\rm end} (u)]$
(if ${\rm beg} (v)={\rm end} (u)+1$, the intersection of $u$ and $v$
is assumed to be an empty word). The length of the intersection of the
overlapped factors $u$ and $v$ is called the {\it overlap} of $u$ and~$v$.
The {\it union} of the overlapped factors $u$ and $v$ is the factor $w[i..j]$
where $i= \min({\rm beg} (u), {\rm beg} (v))$, $j=\max({\rm end} (u), {\rm end} (v))$.
If some word $u$ is equal to a factor~$v$ of~$w$ then $v$ is called {\it an occurrence} 
of~$u$ in~$w$.

A positive integer $p$ is called  a {\it period} of~$w$ if $w[i]=w[i+p]$ 
for each $i=1,\ldots ,n-p$. We denote by $p(w)$ the minimal period of~$w$
and by $e(w)$ the ratio $|w|/p(w)$ which is called the {\it exponent} of~$w$.
Farther we use the following well-known fact which is usually called 
{\it the periodicity lemma}.
\begin{lemma}
If a word $w$ has two periods $p,q$, and $|w|\ge p+q$,
then $\gcd (p,q)$ is also a period of~$w$.
\label{perilemma}
\end{lemma}
The periodicity lemma is actually a weaker version of
the Fine and Wilf theorem (see~\cite{FineWilf, ChoffKarh97}).
Using the periodicity lemma, it is easy to obtain
\begin{proposition}
Let $q$ be a period of a word~$w$ such that $|w|\ge 2q$.
Then $q$ is divisible by $p(w)$.
\label{repper}
\end{proposition}
We will also use the following evident fact.
\begin{proposition}
If two overlapped factors of a word have the same period~$p$ and 
the overlap of these factors is not less than $p$ then $p$ is a period
of the union of these factors.
\label{unionper}
\end{proposition}

A word is called {\it primitive} if its exponent is not an integer greater
than~1. For primitive words the following well-known fact takes place (see, e.g.,~\cite{algonstr}).
\begin{lemma}[primitivity lemma]
If $u$ is a primitive word, then the word $uu$ has no occurrences of~$u$
which are neither prefix nor suffix of~$uu$.
\label{onprimit}
\end{lemma}

Words~$r$ such that $e(r)\ge 2$ are called {\it repetitions}. A repetition in a word is called 
{\it maximal} if this repetition cannot be extended to the left nor to the right in the word 
by at least one letter while preserving its minimal period. More precisely, a repetition 
$r$ in a word~$w$ is called {\it maximal} if it satisfies the following {\it maximality conditions}:
\begin{enumerate}
\item if ${\rm beg}(r)>1$, then $w[{\rm beg}(r)-1]\neq w[{\rm beg}(r)+p(r)-1]$,
\item if ${\rm end}(r)<n$, then $w[{\rm end}(r)-p(r)+1]\neq w[{\rm end}(r)+1]$.
\end{enumerate}
For example, word $\texttt{ababaabaaababab}$ has 6 maximal repetitions: $\texttt{ababa}$,
$\texttt{abaabaa}$, $\texttt{aabaaaba}$, $\texttt{aa}$, $\texttt{aaa}$ and $\texttt{ababab}$. 
Maximal repetitions are usually called {\it runs} in the literature. By ${\cal R}(w)$ we 
denote the set of all maximal repetitions in the word~$w$. For any natural~$n$ we define also
$R(n)=\max_{|w|=n} |{\cal R}(w)|$ and $E(n)=\max_{|w|=n} \sum_{r\in {\cal R}(w)} e(r)$.

The possible number of maximal repetitions was actively investigated in the literature.
In~\cite{KK00} the linear upper bound $E(n)=O(n)$ is proved which implies obviously that
$R(n)=O(n)$. Due to a series of papers (see. e.~g.~\cite{CrochIlieTinta}) more precise 
upper bounds on $E(n)$ and $R(n)$ have been obtained. A breakthrough in this direction was made 
in~\cite{RunsTheor} where the bounds $R(n)<n$, $E(n)<3n$ are proved. To our knowledge, the 
best upper bound $|{\cal R}(w)|\le\frac{183}{193}n$ for binary words~$w$ of length~$n$ is shown in~\cite{Holub17}
and the best lower bounds $2.035 n$ on $E(n)$ and $0.9445757n$ on $R(n)$ are obtained 
respectively in~\cite{Crochetal11} and~\cite{Simpson}. Some results on the average number
of runs in arbitrary words are obtained in~\cite{PugSimp08,Christodoul14}.

In the paper we consider words over (polynomially bounded) integer alphabet,
i.e. words over alphabet which consists of nonnegative integers bounded by some 
polynomial of the length of words. So farther by~$w$ we will mean an arbitrary 
word of length~$n$ over integer alphabet.

The problem of finding of all maximal repetitions in words was also actively investigated 
in the literature. An $O(n)$-time algorithm for finding all runs in a word of length~$n$ 
was proposed in~\cite{KK00} for the case of constant-size alphabet. This result was
generalized to the case of words over integer alphabet in~\cite{CrochIlieSmyth}.
Another $O(n)$-time algorithm for the case of integer alphabet, based on a different approach, 
has been proposed in~\cite{RunsTheor}. Algorithm for solving the problem
in the case of an unbounded linearly-ordered alphabet was proposed in~\cite{Kosolobov16}.
This algorithm was improved in~\cite{Gawrychowski16,Crochemore16}. Finally, a linear time algorithm for this
case was proposed in~\cite{ElFisch21}. The obtained results can be summarized in the following
two theorems.
\begin{theorem}
The number of maximal repetitions in~$w$ is $O(n)$, and all these repetitions with their
minimal periods can be computed in $O(n)$ time.
\label{numtimerun}
\end{theorem}
\begin{theorem}
The sum $\sum_{r\in {\cal R}(w)} e(r)$ of exponents of all maximal repetitions in~$w$ is $O(n)$.
\label{sumexp}
\end{theorem}

Let $r$ be a repetition in~$w$. We call any factor of~$w$ which has the length $p(r)$ 
and is contained in~$r$ a {\it cyclic root} of~$r$. The cyclic root which is the prefix
of~$r$ is called {\it prefix} cyclic root of~$r$. It follows from the minimality of the 
period $p(r)$ that any cyclic root of~$r$ is a primitive word. So the following proposition 
can be easily obtained from Lemma~\ref{onprimit}.
\begin{proposition}
Two cyclic roots $x', x''$ of a repetition~$r$ are equal if and only if 
${\rm beg}(x')\equiv {\rm beg}(x'')\pmod{p(r)}$ (${\rm end}(x')\equiv {\rm end}(x'')\pmod{p(r)}$).
\label{cyceqv}
\end{proposition}
Since all roots of any repetition are primitive, any repetition~$r$ has $p(r)$ different
cyclic roots which are cyclic rotations of each other. The lexicographically minimal root
among these cyclic roots is called {\it Lyndon root} of the repetition~$r$. Let $x$ be the
leftmost occurrence of the Lyndon root in the repetition~$r$. Then the difference
${\rm beg}(x)-{\rm beg}(r)$ is denoted by $a(r)$. Two repetitions with the same minimal 
period are called {\it repetitions with the same cyclic roots} if they have
the same set of distinct cyclic roots. Note that repetitions with the same cyclic roots
has the same Lyndon root.

Let $r', r''$ be maximal repetitions with the same cyclic roots, $p$ be the minimal period
of $r'$ and~$r''$, and $x', x''$ be cyclic roots of repetitions $r', r''$ respectively.
Note that ${\rm beg}(r')+a(r')$ and ${\rm beg}(r'')+a(r'')$ are the starting positions of 
the leftmost Lyndon roots of repetitions $r', r''$ respectively. Denote by $\delta'$ ($\delta''$)
the residue of ${\rm beg}(x')-({\rm beg}(r')+a(r'))$ (${\rm beg}(x'')-({\rm beg}(r'')+a(r''))$) 
in modulo~$p$. Using Proposition~\ref{cyceqv}, it is easy to see that $x'=x''$ if and only if
$\delta'=\delta''$. Thus, we obtain the following fact.
\begin{proposition}
Let $r', r''$ be maximal repetitions with the same cyclic roots, $p$ be the minimal period
of $r'$ and~$r''$, and $x', x''$ be cyclic roots of repetitions $r', r''$ respectively.
Then $x'=x''$ if and only if 
$$
{\rm beg}(x')-({\rm beg}(r')+a(r'))\equiv {\rm beg}(x'')-({\rm beg}(r'')+a(r''))\pmod{p}.
$$
\label{difcyceqv}
\end{proposition}

Farther we will use double-linked lists of all maximal repetitions with the same cyclic roots 
in the order of increasing of their starting positions. According to~\cite{Crochemore14}, these lists
can be computed for the wird~$w$ in $O(n)$ time. It is also shown in~\cite{Crochemore14} that for all 
maximal repetitions~$r$ in~$w$ the values $a(r)$ cab be computed in $O(n)$ total time.

We will also use the following facts for overlaps of maximal
repetitions (see, e.g.~\cite{Kolpakov12}).
\begin{proposition}
The overlap of any two different maximal repetitions 
with the same minimal period~$p$ is smaller than~$p$.
\label{repoverlap0}
\end{proposition}
\begin{proposition}
The overlap of any two different maximal repetitions 
$r'$ and $r''$ is smaller than $p(r')+p(r'')$.
\label{repoverlap1}
\end{proposition}

A natural generalization of repetitions are factors with exponents strictly less than~2.
We will call such factors {\it subrepetitions}. More precisely, a factor~$r$ is called a
{\it subrepetition} if $1<e(r)<2$. Analogously to repetitions, a subrepetition~$r$ in~$w$
is called maximal if $r$ satisfies the maximality conditions, i.e. if $r$  cannot be extended 
to the left nor to the right in~$w$ by at least one letter while preserving its minimal period.
For any $\delta$ such that such that $0<\delta <1$\footnote{In the paper for convenience we assume 
actually that $\delta <1-\varepsilon$ for some fixed~$\varepsilon$.} a subrepetition~$r$ is called 
({\it $\delta$-subrepetition}) if $e(r)\ge 1+\delta$. It is shown below that the number
of maximal $\delta$-subrepetitions in a word of length~$n$ is $O(n/\delta)$. In this paper 
the following problem is investigated.
\begin{problem}
For a given value~$\delta$ find in a given word~$w$ of length~$n$ all maximal 
$\delta$-subrepetions.
\label{firstprob}
\end{problem}

Before in~\cite{forJDA} two algorithms for solving of Problem~\ref{firstprob} was proposed:
the first algorithm has $O(\frac{n\log\log n}{\delta^2})$ time complexity and the second 
algorithm has $O(n\log n+\frac{n}{\delta^2}\log \frac{1}{\delta})$ expected time complexity.
In~\cite{Kociumaka15} the expected time of the second algorithm was improved to the linear bound 
$O(\frac{n}{\delta^2}\log \frac{1}{\delta})$. Using the results of~\cite{Crochemore19,GawrychowskiIIK18}, this
time can be farther improved to $O(\frac{n}{\delta}\log \frac{1}{\delta})$.
In~\cite{BadkobehCrochToop12}, it is shown that all subrepetitions with the largest exponent (over all subrepetitions) 
in an overlap-free string can be found in O(n) time for a constant-size alphabet.
In this paper we propose an alternative deterministic algorithm for solving of Problem~\ref{firstprob} 
in $O(\frac{n}{\delta}\log\frac{1}{\delta})$ time.

\section{Repeats}

Another regular structures in a word which are closely related to repetitions and
subrepetitions are {\it repeats}. In general case, a repeat~$\sigma$ in the word~$w$ is 
a pair $u', u''$ of equal factors of the word~$w$ where ${\rm beg}(u')<{\rm beg}(u'')$. 
The factors $u', u''$ are called {\it copies} of the repeat~$\sigma$, in particular,
$u'$ is called {\it the left copy} of~$\sigma$ and $u'$ is called {\it the right copy} 
of~$\sigma$. The length of copies of~$\sigma$ is denoted by $c(\sigma)$. The difference
${\rm beg}(u'')-{\rm beg}(u')$ is called the {\it period} of the repeat~$\sigma$ and
is denoted by $p(\sigma)$. The minimal factor $w[{\rm beg}(u')..{\rm end}(u'')]$
containing the both copies $u', u''$ of~$\sigma$ will be denoted by ${\rm fact}(\sigma)$.
Note that for different repeats $\sigma'$ and $\sigma''$ we can have actually
${\rm fact}(\sigma')={\rm fact}(\sigma'')$. Note also that $p(\sigma)$ is a period 
of ${\rm fact}(\sigma)$, but the minimal period of ${\rm fact}(\sigma)$
can be less than $p(\sigma)$. By the starting position (the ending position) 
of~$\sigma$ we will mean the starting position (the ending position) of ${\rm fact}(\sigma)$. 
We will say that a maximal repeat $\sigma$ is contained in some factor $w'$ of the word 
if the factor ${\rm fact}(\sigma)$ is contained in $w'$. A repeat is called {\it overlapped} 
if its copies are overlapped factors, otherwise the repeat is called {\it gapped}. 
In other words, the repeat~$\sigma$ is gapped if ${\rm fact}(\sigma)$ can be represented
in the form $u'vu''$ where $v$ is a nonempty word called the {\it gap} of the repeat~$\sigma$.
For any $\alpha>1$ a gapped repeat~$\sigma$ is called {\it $\alpha$-gapped} if 
$p(\sigma)\le\alpha c(\sigma)$.

Analogously to repetitions and subrepetitions, we can also introduce a notion of maximal repeats. 
A repeat~$\sigma$ with left and right copies $u', u''$ in~$w$ is called {\it maximal} if
it satisfies the following conditions:
\begin{enumerate}
\item if ${\rm beg}(u')>1$, then $w[{\rm beg}(u')-1]\neq w[{\rm beg}(u'')-1]$,
\item if ${\rm end}(u'')<n$, then $w[{\rm end}(u')+1]\neq w[{\rm end}(u'')+1]$.
\end{enumerate}
In other words, a repeat in a word is called {\it maximal} if its copies cannot be extended to the 
left nor to the right in the word by at least one letter while preserving its period.
Note that any repeat can be extended to uniquely defined maximal repeat with the same period.
In particular, any $\alpha$-gapped repeat can be extended to uniquely defined maximal $\alpha$-gapped 
or overlapped repeat. In~\cite{Crochemore19,GawrychowskiIIK18} the following fact on maximal 
$\alpha$-gapped repeats was obtained.
\begin{theorem}
The number of maximal $\alpha$-gapped repeats in~$w$ is $O(\alpha n)$, and
all these repeats can be computed in $O(\alpha n)$ time. 
\label{numtimereps}
\end{theorem}
In~\cite{GawrychowskiIIK18} the more precise upper bound $18\alpha n$ on the number of maximal $\alpha$-gapped 
repeats in~$w$ was actually obtained. A tighter bound on this number was obtained later in~\cite{IKoppl19}. 
An algorithm which finds in each position of word longest gapped repeats satisfying additional 
restrictions is proposed in~\cite{Dumitran15}.

Let $\sigma$ be an overlapped repeat with with left and right copies $u', u''$ in~$w$.
Note that in this case the period $p(\sigma)$ of ${\rm fact}(\sigma)$ is not greater than
the half of $|{\rm fact}(\sigma)|$, so ${\rm fact}(\sigma)$ is a repetition. Let $p$ be
the minimal period of ${\rm fact}(\sigma)$. By Proposition~\ref{repper} we have that $p$ is a
divisor of $p(\sigma)$. Let ${\rm beg}({\rm fact}(\sigma))={\rm beg}(u')>1$. Since $\sigma$ is 
a maximal repeat, we have that 
$$
w[{\rm beg}({\rm fact}(\sigma))-1]=w[{\rm beg}(u')-1]\neq w[{\rm beg}(u'')-1]=w[{\rm beg}({\rm fact}(\sigma))+p(\sigma)-1].
$$
On the other hand, since the period $p$ of ${\rm fact}(\sigma)$ is a divisor of $p(\sigma)$, we have 
that $w[{\rm beg}({\rm fact}(\sigma))+p-1]=w[{\rm beg}({\rm fact}(\sigma))+p(\sigma)-1]$. Thus, we 
obtain that $w[{\rm beg}({\rm fact}(\sigma))-1]\neq w[{\rm beg}({\rm fact}(\sigma))+p-1]$.
In analogous way we can obtain that if ${\rm end}({\rm fact}(\sigma))<n$ then 
$w[{\rm beg}({\rm fact}(\sigma))+1]\neq w[{\rm beg}({\rm fact}(\sigma))-p+1]$.
Thus, we conclude that ${\rm fact}(\sigma)$ is a maximal repetition whose minimal period
is a divisor of $p(\sigma)$. We will denote this repetition by ${\rm rep}(\sigma)$.
Now let $r$ be a maximal repetition in~$w$. Then we can consider the repeat $\sigma$ with left and right 
copies $w[{\rm beg}(r) .. {\rm end}(r)-p(r)]$ and $w[{\rm beg}(r)+p(r) .. {\rm end}(r)]$.
It is easy to note that $\sigma$ is a maximal overlapped repeat such that $p(\sigma)=p(r)$.
We will call the repeat~$\sigma$ the {\it principal repeat} of the repetition~$r$.
Principal repeats of maximal repetitions will be also called {\it reprincipal repeats}.
Note that any reprincipal repeat~$\sigma$ is the principal repeats of the repetition
${\rm rep}(\sigma)$, so for any maximal repetition we have that this repetition and 
the principal repeat of this repetition are uniquely defined by each other. Therefore, we
have one-to-one correspondence between all maximal repetitions and all reprincipal repeats
in a word. Thus, in any word the number of reprincipal repeats is equal to the number of 
maximal repetitions. We have the following fact for reprincipal repeats.
\begin{proposition}
The number of reprincipal repeats in~$w$ is $O(n)$, and all these repeats can be computed in $O(n)$ time.
\label{findprincip}
\end{proposition}
{\bf Proof.} Recall that the number of reprincipal repeats in~$w$ is equal to the number of maximal repetitions
in~$w$, so, by Theorem~\ref{numtimerun}, this number is $O(n)$. To compute all reprincipal repeats in~$w$,
first we find all maximal repetitions in~$w$. By Theorem~\ref{numtimerun}, it can be done in time $O(n)$.
Then for each found maximal repetition~$r$ we compute the principal repeat of~$r$ as repeat $(u', u'')$ where
${\rm beg}(u')={\rm beg}(r)$, ${\rm end}(u')={\rm end}(r)-p(r)$, ${\rm beg}(u'')={\rm beg}(r)+p(r)$, 
${\rm end}(u'')={\rm end}(r)$. It can be computed in constant time. Therefore, the total time of computing of
all reprincipal repeats from found maximal repetitions is proportional to the number of these repetitions, so
this time is $O(n)$. Thus, the total time of computing of all reprincipal repeats in~$w$ is $O(n)$. \qed

\begin{figure}[!ht]
\centerline{
\includegraphics[height=30mm,width=60mm]{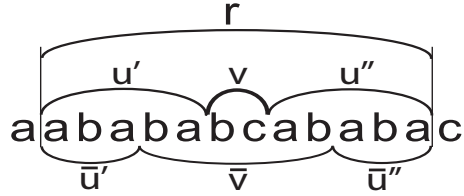}
}
\caption{Principal and non principal repeats.}
\label{prinrep}
\end{figure}

Now let $r$ be a maximal $\delta$-subrepetition in~$w$. Then we can consider a gapped repeat~$\sigma$ with
the left copy $w[{\rm beg}(r) .. {\rm end}(r)-p(r)]$ and the right copy $w[{\rm beg}(r)+p(r) .. {\rm end}(r)]$.
It is easy to see that $\sigma$ is a maximal gapped repeat with the period $p(\sigma)=p(r)$. Moreover, we have
$$
c(\sigma)=|r|-p(r)\ge (1+\delta)p(r)-p(r)=\delta p(r)=\delta p(\sigma),
$$
so $p(\sigma)\le c(\sigma)/\delta$. Thus, $\sigma$ is actually a maximal $\frac{1}{\delta}$-gapped repeat.
We call the repeat~$\sigma$ the {\it principal repeat} of the subrepetition~$r$. Thus, for any maximal 
$\delta$-subrepetition there exists the principal repeat of this subrepetition. On the other hand, 
a maximal gapped repeat may not be the principal repeat of any maximal subrepetition.  For example, in the 
word shown in Fig.~\ref{prinrep} we can consider the maximal subrepetition~$r$ with the minimal period~7.
Note that the gapped repeat $u'vu''$ is the principal repeat of~$r$ while the gapped repeat 
$\overline{u}'\overline{v}\;\overline{u}''$ is not the principal repeat of~$r$, so the repeat 
$\overline{u}'\overline{v}\;\overline{u}''$ is not the principal repeat of any maximal subrepetition.
Thus, the repeat $u'vu''$ is a principal repeat and the repeat $\overline{u}'\overline{v}\;\overline{u}''$
is not a principal repeat. Note that for any maximal subrepetition we have that this subrepetition and 
the principal repeat of this subrepetition are uniquely defined by each other, and we have one-to-one 
correspondence between all maximal $\delta$-subrepetitions and all principal maximal $\frac{1}{\delta}$-gapped 
repeats in a word. Thus, Problem~\ref{firstprob} can be reformulated in the following way.
\begin{problem}
For a given value~$\delta$ find in a given word~$w$ of length~$n$ all principal maximal $1/\delta$-gapped repeats.
\label{secondprob}
\end{problem}
We obtain also that in any word the number of maximal $\delta$-subrepetitions is not greater than
the number of maximal $\frac{1}{\delta}$-gapped repeats. Thus, Theorem~\ref{numtimereps} implies
the following upper bound on the number of maximal $\delta$-subrepetitions in a word.
\begin{proposition}
The number of maximal $\delta$-subrepetitions in~$w$ is $O(n/\delta)$.
\label{numdelsubs}
\end{proposition}

Not also that for any principal overlapped or gapped repeats we have the following obvious fact.
\begin{proposition}
A maximal repeat $\sigma$ is principal if and only if the minimal period of 
${\rm fact}(\sigma)$ is equal to $p(\sigma)$.
\label{onprincip}
\end{proposition}

Let $M$ be a set of maximal repeats in the word~$w$. Note that a maximal repeat 
is uniquely defined by its starting position together with its period. So we can 
represent the set~$M$ by lists $ML_t$ for $t=1, 2,\ldots , n$ where $ML_t$ 
is the list containing all these repeats with the starting position~$t$ 
in the order of increasing of their periods. Using the bucket sorting, all 
the lists $ML_t$ can be computed from the set~$M$ in total $O(n+|M|)$ time. 
Moreover, all the lists $ML_t$ can be traversed in total $O(n+|M|)$ time.
We will call the lists $ML_t$ {\it start position lists} for the set~$M$.

%%%%%%%%%%%%%%%%%%%%%%%%%%%%%%%%%%%%%%%%%%%%%%%%%%%%%%%%%%%%%%%%%%%%%%%%%%%%%%%%%%%%%%%%%%%%%%%%%%%%%%%%%%%
\section{Covering of repeats}
%%%%%%%%%%%%%%%%%%%%%%%%%%%%%%%%%%%%%%%%%%%%%%%%%%%%%%%%%%%%%%%%%%%%%%%%%%%%%%%%%%%%%%%%%%%%%%%%%%%%%%%%%%%

We say that a maximal repeat~$\sigma$ is {\it covered} by another maximal 
repeat~$\sigma'$ if ${\rm fact}(\sigma)$ is contained in ${\rm fact}(\sigma')$ and $p(\sigma')<p(\sigma)$. 
Note that the introduced notion of covering of repeats satisfies the transitivity 
property: if some maximal repeat $\sigma'$ is covered by some maximal repeat $\sigma''$
and the maximal repeat $\sigma''$ is covered by some maximal repeat $\sigma'''$ then
$\sigma'$ is covered by $\sigma'''$. The following auxiliary facts can be easily checked.
\begin{proposition}
A maximal $\alpha$-gapped repeat can be covered only by $\alpha$-gapped
or overlapped repeats.
\label{onlycover}
\end{proposition}
\begin{proposition}
If a maximal repeat~$\sigma$ is covered by a maximal repeat~$\sigma'$ then
the left copy of~$\sigma$ is contained in the left copy of~$\sigma'$ and
the right copy of~$\sigma$ is contained in the right copy of~$\sigma'$.
\label{copycover}
\end{proposition}

In analogous way, a maximal repeat~$\sigma$ is covered by a maximal
repetition~$r$ if ${\rm fact}(\sigma)$ is contained in~$r$ and $p(r)<p(\sigma)$.
Note that the repeat~$\sigma$ is covered by~$r$ if and only if $\sigma$ 
is covered by the principal repeat of~$r$. Note also that any maximal
overlapped repeat coincides as factor with some maximal repetition
whose period is a divisor of the period of this repeat. So any
maximal repeat covered by a maximal overlapped repeat is covered also
by the maximal repetition coinciding with this overlapped repeat.
Thus we have the following fact.
\begin{proposition}
Any maximal repeat covered by a maximal overlapped repeat is covered also
by the principal repeat of some maximal repetition.
\label{overcover}
\end{proposition}

Let a maximal repeat $\sigma$ be covered by a maximal repeat $\sigma'$.
Then the factor ${\rm fact}(\sigma)$ has the period $p(\sigma')<p(\sigma)$, so, 
by Proposition~\ref{onprincip}, the repeat $\sigma$ is not principal. Hence 
principal repeats are not covered by other maximal repeats. Now let a maximal 
repeat $\sigma$ be not principal, i.e. $\sigma$ is contained in some repetition 
or subrepetition~$r$ such that $p(r)<p(\sigma)$. In this case $\sigma$ is covered 
by the principal repeat of~$r$. Therefore, if $\sigma$ is not covered by any other 
maximal repeat then $\sigma$ is principal. Thus, we obtain the following fact.
\begin{proposition}
A maximal repeat is principal if and only if it is not covered by any other maximal repeat.
\label{princrit}
\end{proposition}

Using Propositions \ref{princrit} and~\ref{onlycover}, Problem~\ref{secondprob} can be reformulated 
in the following way.
\begin{problem}
For a given value~$\delta$ find in a given word~$w$ of length~$n$ all maximal $1/\delta$-gapped repeats 
which are not covered by other maximal repeats.
\label{thirdprob}
\end{problem}

%%%%%%%%%%%%%%%%%%%%%%%%%%%%%%%%%%%%%%%%%%%%%%%%%%%%%%%%%%%%%%%%%%%%%%%%%%%%%%%%%%%%%%%%%%%%%%%%%%%%%%%%%%%
\section{Periodic and generated repeats}
%%%%%%%%%%%%%%%%%%%%%%%%%%%%%%%%%%%%%%%%%%%%%%%%%%%%%%%%%%%%%%%%%%%%%%%%%%%%%%%%%%%%%%%%%%%%%%%%%%%%%%%%%%%

Repeat~$\sigma$ is called {\it periodic} if the copies of~$\sigma$ are repetitions
with minimal period not greater than $c(\sigma )/3$, otherwise $\sigma$ is called 
{\it nonperiodic}. Let $\sigma$ be a maximal periodic repeat with a left copy~$u'$
and a right copy $u''$. Since $u'$ and $u''$ are repetitions, these repetitions
can be extended to some maximal repetitions $r'$ and~$r''$ with the same cyclic
roots. Let $r'$ and~$r''$ be different maximal repetitions. Then we say that
$\sigma$ is {\it represented by the pair of maximal repetitions} $r', r''$ or,
more briefly, $\sigma$ is {\it birepresented}. The pair of maximal repetitions $r', r''$
will be called {\it left} if $|r'|\le |r''|$, otherwise it is called {\it right}.
We will also call the repeat~$\sigma$ {\it left (right) birepresented} if $\sigma$
is represented by a left (right) pair of maximal repetitions. If the maximal 
repetitions $r'$ and~$r''$ are the same repetition~$r$, we say that $\sigma$ is 
{\it represented by the maximal repetition}~$r$. 

A maximal repeat~$\sigma$ is called to be {\it generated} by a maximal repetition~$r$ 
if ${\rm beg}(\sigma)={\rm beg}(r)$, ${\rm end}(\sigma)={\rm end}(r)$, and $p(\sigma)$ 
is divisible by $p(r)$. For maximal periodic repeats represented by maximal repetitions 
we have the following fact.
\begin{proposition}
Any maximal periodic repeat represented by a maximal repetition is generated by
this maximal repetition.
\label{repmaxrep}
\end{proposition}
{\bf Proof.} Let $\sigma$ be a maximal periodic repeat represented by a maximal repetition~$r$,
and $u'$, $u''$ be respectively the left and right copies of~$\sigma$. Note that, since $\sigma$ 
is periodic, the length of $u'$ and $u''$ is greater than $p(r)$. Denote by $x'$, $x''$ 
the cyclic roots of~$r$ which are prefixes of $u'$ and $u''$ respectively. Since $x'=x''$,
by Proposition~\ref{cyceqv}, we have ${\rm beg}(x')\equiv {\rm beg}(x'')\pmod{p(r)}$, so
${\rm beg}(u')\equiv {\rm beg}(u'')\pmod{p(r)}$. Thus, $p(\sigma)={\rm beg}(u'')-{\rm beg}(u')$
is divisible by $p(r)$. Let ${\rm beg}(\sigma)={\rm beg}(u')>{\rm beg}(r)$. Then the both
symbols $w[{\rm beg}(u')-1]$ and $w[{\rm beg}(u'')-1]$ are contained in~$r$ and the difference
between the positions of these symbols is divisible by $p(r)$. So $w[{\rm beg}(u')-1]=w[{\rm beg}(u'')-1]$
which contradicts that $\sigma$ is a maximal repeat. Thus, ${\rm beg}(\sigma)={\rm beg}(r)$.
In an analogous way, we have that ${\rm end}(\sigma)={\rm end}(r)$. Thus, $\sigma$ is generated by~$r$. \qed

Note that any maximal repetition~$r$ generates no more than $e(r)/2$ gapped repeats and, knowing values
${\rm beg}(r)$, ${\rm end}(r)$ and $p(r)$, we can compute all these repeats in $O(e(r))$ time.
We can check each of these repeats in constant time if this repeat is $\alpha$-gapped.
Thus, we can compute all maximal $\alpha$-gapped repeats generated by~$r$ in $O(e(r))$ time.
So we have the following simple procedure of computing of all maximal $\alpha$-gapped repeats generated 
by maximal repetitions in~$w$. First we find all maximal repetitions in~$w$. According to Theorem~\ref{numtimerun}, 
it can be done in $O(n)$ time, and the total number of these repetitions is $O(n)$. 
Then for each found repetition~$r$ we compute all maximal $\alpha$-gapped repeats generated by~$r$ 
in $O(e(r))$ time. The total time of these procedure is $O(\sum_{r\in {\cal R}(w)} e(r))$, so, 
by Theorem~\ref{sumexp}, this time is $O(n)$. Note also that, by Theorem~\ref{sumexp}, the number of computed 
repeats is $O(n)$. Thus, we obtain the following fact.
\begin{proposition}
The number of maximal $\alpha$-gapped repeats generated by maximal repetitions in~$w$ is $O(n)$, and
all these repeats can be computed in $O(n)$ time. 
\label{findgenreps}
\end{proposition}

%%%%%%%%%%%%%%%%%%%%%%%%%%%%%%%%%%%%%%%%%%%%%%%%%%%%%%%%%%%%%%%%%%%%%%%%%%%%%%%%%%%%%%%%%%%%%%%%%%%%%%%%%%%
\section{Birepresented gapped repeats}
%%%%%%%%%%%%%%%%%%%%%%%%%%%%%%%%%%%%%%%%%%%%%%%%%%%%%%%%%%%%%%%%%%%%%%%%%%%%%%%%%%%%%%%%%%%%%%%%%%%%%%%%%%%

Let the maximal repeat~$\sigma$ with left and right copies $u'$ and $u''$ be represented by a left pair 
of maximal repetitions $r', r''$ with the same cyclic roots, and $p$ be the minimal period of $r'$ and $r''$. 
Assume that ${\rm beg}(u')>{\rm beg}(r')$ and ${\rm beg}(u'')>{\rm beg}(r'')$. Then $w[{\rm beg}(u')-1]$ 
and $w[{\rm beg}(u'')-1]$ are contained in $r'$ and $r''$ respectively, so
\begin{eqnarray*}
u'[p]&=&w[{\rm beg}(u')-1+p]=w[{\rm beg}(u')-1],\\
u''[p]&=&w[{\rm beg}(u'')-1+p]=w[{\rm beg}(u'')-1].
\end{eqnarray*}
Thus, from $u'[p]=u''[p]$ we have $w[{\rm beg}(u')-1]=w[{\rm beg}(u'')-1]$ which contradicts 
that $\sigma$ is maximal. Therefore, we have ${\rm beg}(u')={\rm beg}(r')$ or ${\rm beg}(u'')={\rm beg}(r'')$.
Analogously, it can be shown that ${\rm end}(u')={\rm end}(r')$ or ${\rm end}(u'')={\rm end}(r'')$.
Note that, since $|r'|\le |r''|$, the case of ${\rm beg}(u'')={\rm beg}(r'')$ and ${\rm end}(u'')={\rm end}(r'')$
implies that ${\rm beg}(u')={\rm beg}(r')$ and ${\rm end}(u')={\rm end}(r')$. Taking it into account,
generally, we can consider separately the three following cases:\\
1. ${\rm beg}(u')>{\rm beg}(r')$, ${\rm end}(u')={\rm end}(r')$ and ${\rm beg}(u'')={\rm beg}(r'')$;\\
2. ${\rm beg}(u')={\rm beg}(r')$, ${\rm end}(u')={\rm end}(r')$ and ${\rm end}(u'')<{\rm end}(r'')$;\\
3. ${\rm beg}(u')={\rm beg}(r')$ and ${\rm end}(u'')={\rm end}(r'')$.\\
We will call the repeat~$\sigma$ {\it repeat of first type} in the case~1, 
{\it repeat of second type} in the case~2, and {\it repeat of third type} in the case~3.

First consider the case~1. Let $\sigma'$, $\sigma''$ be two different maximal repeats of first type
represented by the left pair of maximal repetitions $r', r''$. If ${\rm beg}(\sigma')={\rm beg}(\sigma'')$
then it is easy to note that $\sigma'$ and $\sigma''$ are the same repeat, so ${\rm beg}(\sigma')\neq {\rm beg}(\sigma'')$.
Let $x'$, $x''$ be the prefixes of length~$p$ in the left copies of repeats $\sigma'$ and $\sigma''$ respectively.
Note that $x'$, $x''$ are cyclic roots of $r'$ which are equal to the prefix cyclic root of~$r''$, so $x'=x''$.
Thus, by Proposition~\ref{cyceqv}, we have that the difference ${\rm beg}(x')-{\rm beg}(x'')=
{\rm beg}(\sigma')-{\rm beg}(\sigma'')$ is divisible by~$p$. Moreover, by definition of repeats
of first type, we have that both values ${\rm beg}(\sigma'), {\rm beg}(\sigma'')$ are in the segment
$$
{\rm beg}(r')<{\rm beg}(\sigma'), {\rm beg}(\sigma'')\le {\rm end}(r')-3p+1.
$$
Thus, the starting positions of all maximal repeats of first type represented by the pair $r', r''$
form in the segment $[{\rm beg}(r')+1; {\rm end}(r')-3p+1]$ an arithmetic progression of numbers with 
the step~$p$. Let $t'$ be the maximal number in this progression and $k'$ be the number of all maximal 
repeats of first type represented by the pair $r', r''$. Then we can consider the numbers
$t', t'-p, t'-2p,\ldots , t'-(k'-1)p$ of this progression in descending order as the starting positions
of the corresponding repeats. In this way, we consider the set of all maximal repeats of first type 
represented by the pair $r', r''$ as $\{\sigma'_1, \sigma'_2,\ldots , \sigma'_{k'}\}$ where
${\rm beg}(\sigma'_j)=t'-(j-1)p$ for $j=1, 2,\ldots , k'$. Note that 
${\rm beg}(\sigma'_j)={\rm beg}(\sigma'_1)-(j-1)p$, $p(\sigma'_j)=p(\sigma'_1)+(j-1)p$ and
${\rm end}(\sigma'_j)={\rm end}(\sigma'_1)+(j-1)p$ for $j=1, 2,\ldots , k'$.

Now consider the case 2. Let $\sigma'$, $\sigma''$ be two different maximal repeats of second type
represented by the left pair of maximal repetitions $r', r''$, and $\hat u'$, $\hat u''$ be the right copies
of repeats $\sigma'$, $\sigma''$ respectively. If ${\rm beg}(\hat u')={\rm beg}(\hat u'')$ then $\sigma'$ and 
$\sigma''$ are the same repeat, so ${\rm beg}(\hat u')\neq {\rm beg}(\hat u'')$. Let $x'$, $x''$ be 
the prefixes of length~$p$ in $\sigma'$ and $\sigma''$ respectively. Note that $x'$, $x''$ are cyclic roots 
of $r''$ which are equal to the prefix cyclic root of~$r'$, so $x'=x''$. Thus, by Proposition~\ref{cyceqv}, 
we have that the difference ${\rm beg}(\hat u')-{\rm beg}(\hat u'')$ is divisible by~$p$. Moreover, 
by definition of repeats of second type, we have that both values ${\rm beg}(\hat u'), {\rm beg}(\hat u'')$
are in the segment
$$
{\rm beg}(r'')\le {\rm beg}(\hat u'), {\rm beg}(\hat u'')\le {\rm end}(r'')-|r'|.
$$
Thus, the starting positions of right copies of all maximal repeats of second type represented by the 
pair $r', r''$ form in the segment $[{\rm beg}(r''); {\rm end}(r'')-|r'|]$ an arithmetic progression of 
numbers with the step~$p$. Let $t''$ be the minimal number in this progression and $k''$ be the number 
of all maximal repeats of second type represented by the pair $r', r''$. Then we can consider the numbers
$t'', t''+p, t''+2p,\ldots , t''+(k''-1)p$ of this progression in ascending order as the starting positions
of the right copies of the corresponding repeats. In this way, we consider the set of all maximal repeats of 
second type  represented by the pair $r', r''$ as $\{\sigma''_1, \sigma''_2,\ldots , \sigma''_{k''}\}$ where
the starting position of the right copy of $\sigma''_j$ is $t''+(j-1)p$ for $j=1, 2,\ldots , k''$. Note that
${\rm beg}(\sigma''_j)={\rm beg}(r')$, $p(\sigma''_j)=p(\sigma''_1)+(j-1)p$ and 
${\rm end}(\sigma''_j)={\rm end}(\sigma'_1)+(j-1)p$ for $j=1, 2,\ldots , k''$.

Finally consider the case 3. Let $\sigma'$, $\sigma''$ be two different maximal repeats of third type
represented by the left pair of maximal repetitions $r', r''$, and $\hat u'$, $\hat u''$ be the right 
copies of repeats $\sigma'$, $\sigma''$ respectively. Analogously to case 2, it can be shown that 
${\rm beg}(\hat u')\neq {\rm beg}(\hat u'')$ and the difference ${\rm beg}(\hat u')-{\rm beg}(\hat u'')$ 
is divisible by~$p$. Thus, the starting positions of right copies of all maximal repeats of third type 
represented by the pair $r', r''$ form in the segment $[{\rm end}(r'')-|r'|+1; {\rm end}(r'')-3p+1]$ an 
arithmetic progression of numbers with the step~$p$. Let $t'''$ be the minimal number in this progression 
and $k'''$ be the number of all maximal repeats of third type represented by the pair $r', r''$.
Then we can consider the numbers $t''', t'''+p, t'''+2p,\ldots , t'''+(k'''-1)p$ of this progression in 
ascending order as the starting positions of the right copies of the corresponding repeats. In this way, 
we consider the set of all maximal repeats of third type  represented by the pair $r', r''$ as 
$\{\sigma'''_1, \sigma'''_2,\ldots , \sigma'''_{k'''}\}$ where the starting position of the right copy 
of $\sigma'''_j$ is $t'''+(j-1)p$ for $j=1, 2,\ldots , k'''$. Note that ${\rm beg}(\sigma'''_j)={\rm beg}(r')$, 
$p(\sigma'''_j)=p(\sigma'''_1)+(j-1)p$ and ${\rm end}(\sigma'''_j)={\rm end}(r'')$ for $j=1, 2,\ldots , k'''$.
The repeat $\sigma'''_1$ will be called {\it dominating}, and other repeats $\{\sigma'''_2,\ldots , \sigma'''_{k'''}\}$
will be called {\it nondominating}.

Consider additionally the repeats $\sigma'_{k'}$ and $\sigma''_1$. Let $u'_0$ and $u''_0$ be left
and right copies of $\sigma'_{k'}$, and $u'_1$ and $u''_1$ be left and right copies of $\sigma''_1$.
Consider in~$r'$ the prefix $x'$ of length~$p$ which is a cyclic root of $r'$. Since $x'$ is a prefix 
of $u'_1$, we have also in $u''_1$ the prefix $x''$ of length~$p$ which is a cyclic root of $r''$ and
is equal to~$x'$. Note that $x''$ is a factor of $u''_0$, so we can also consider in $u'_0$ the
factor $x'''$ corresponding to $x''$. Note also that $x'''$ is a cyclic root of $r'$ such that $x'''=x'$.
Moreover, $x'''$ has to be in $r'$ the leftmost cyclic root to the right of~$x'$, so, by Proposition~\ref{cyceqv},
${\rm beg}(x''')={\rm beg}(x')+p$. We have also that ${\rm beg}(x'')={\rm beg}(x')+p(\sigma''_1)$ and
${\rm beg}(x''')={\rm beg}(x'')-p(\sigma'_{k'})$, so ${\rm beg}(x''')={\rm beg}(x')+p(\sigma''_1)-p(\sigma'_{k'})$.
Thus, $p(\sigma''_1)-p(\sigma'_{k'})=p$, and 
$$
{\rm end}(\sigma''_1)={\rm end}(r')+p(\sigma''_1)={\rm end}(r')+p(\sigma'_{k'})+p={\rm end}(\sigma'_{k'})+p.
$$

Consider also the repeats $\sigma''_{k''}$ and $\sigma'''_1$. Let $u'_0$ and $u''_0$ be left
and right copies of $\sigma''_{k''}$, and $u'_1$ and $u''_1$ be left and right copies of $\sigma'''_1$.
Consider in~$r'$ the prefix $x'$ of length~$p$ which is a cyclic root of $r'$. Since $x'$ is a prefix 
of $u'_0$, we have in $u''_0$ the prefix $x''$ of length~$p$ which is a cyclic root of $r''$ and
is equal to~$x'$. Since $x'$ is a prefix of $u'_1$, we have also in $u''_1$ the prefix $x'''$ of length~$p$ 
which is a cyclic root of $r''$ and is equal to~$x'$. Thus, $x''$ and $x'''$ are two equal cyclic 
roots of $r''$. Note that $x'''$ has to be in $r'$ the leftmost cyclic root to the right of~$x''$,
so, by Proposition~\ref{cyceqv}, ${\rm beg}(x''')={\rm beg}(x'')+p$. Therefore,
\begin{eqnarray*}
p(\sigma'''_1)&=&{\rm beg}(x''')-{\rm beg}(u'_1)={\rm beg}(x''')-{\rm beg}(u'_0)={\rm beg}(x'')+p-{\rm beg}(u'_0)\\
&=&p(\sigma''_{k''})+p.
\end{eqnarray*}

Now we can join all the repeats represented by the pair of repetitions $r', r''$ into the
sequence of repeats $\Psi =\hat\sigma_1, \hat\sigma_2,\ldots , \hat\sigma_{k'+k''+k'''}$ where
the repeats of first type, the repeats of second type and the repeats of third type are
inserted consecutively, i.e. $\hat\sigma_j=\sigma'_j$ for $j=1, 2,\ldots ,k'$,
$\hat\sigma_{k'+j}=\sigma''_j$ for $j=1, 2,\ldots ,k''$, and $\hat\sigma_{k'+k''+j}=\sigma'''_j$ 
for $j=1, 2,\ldots ,k'''$. From the above observations we have that 
$p(\hat\sigma_{j+1})=p(\hat\sigma_{j})+p$ for $j=1, 2,\ldots , k'+k''+k'''-1$, 
${\rm beg}(\hat\sigma_{j+1})<{\rm beg}(\hat\sigma_{j})$ for $j=1, 2,\ldots , k'$,
${\rm beg}(\hat\sigma_{j})={\rm beg}(r')$ for $j=k'+1, k'+2,\ldots , k'+k''+k'''$,
${\rm end}(\hat\sigma_{j+1})>{\rm end}(\hat\sigma_{j})$ for $j=1, 2,\ldots , k'+k''$,
and ${\rm end}(\hat\sigma_{j})={\rm end}(r'')$ for $j=k'+k''+1, k'+k''+2,\ldots , k'+k''+k'''$.
Note also that ${\rm end}(r'')={\rm end}(\hat\sigma_{k'+k''+1})\le {\rm end}(\hat\sigma_{k'+k''})+p$.
Since $p(\hat\sigma_{j'})<p(\hat\sigma_{j''})$ for $j'<j''$, any repeat $\hat\sigma_{j}$
can not be covered by repeats from~$\Psi$ with greater indexes. Moreover, since 
$|\hat\sigma_{j'}|<|\hat\sigma_{j''}|$ for $j'<j''\le k'+k''+1$, any repeat $\hat\sigma_{j}$ 
for $j\le k'+k''+1$ can not be covered by repeats from~$\Psi$ with smaller indexes.
Thus, any repeat $\hat\sigma_{j}$ for $j\le k'+k''+1$ can not be covered by other repeats
from~$\Psi$. On the other hand, all repeats $\hat\sigma_{j}$ for $j>k'+k''+1$ which are
actually nondominating repeats of third type are covered by the repeat $\hat\sigma_{k'+k''+1}$
which is actually the dominating repeat of third type. Thus we have the following fact.
\begin{proposition}
A maximal repeat~$\sigma$ represented by a left pair of maximal repetitions is covered
by another repeat represented by the same pair of repetitions if and only if $\sigma$
is a nondominating repeat of third type.
\label{prnondleft}
\end{proposition}

From the left pair of maximal repetitions $r', r''$ one can compute effectively all maximal
$\alpha$-gapped periodic repeats represented by the pair $r', r''$.
\begin{lemma}
Let $r', r''$ be a left pair of maximal repetitions with the same cyclic roots such that
$p(r')=p(r'')$, ${\rm beg}(r')$, ${\rm beg}(r'')$, ${\rm end}(r')$, ${\rm end}(r'')$, 
$a(r')$, $a(r'')$ be known. Then all maximal $\alpha$-gapped periodic repeats represented 
by the pair $r', r''$ can be computed in time $O(1+s)$ where $s$ is the number of the 
maximal $\alpha$-gapped periodic repeats represented by the pair $r', r''$.
\label{repreleft}
\end{lemma}

{\bf Proof.} First we compute in constant time sequence~$\Psi$ of repeats where by computing
of~$\Psi$ we mean computing of formulas by which any repeat $\hat\sigma_{j}$ from~$\Psi$
can be computed in constant time. Since any maximal repeat~$\sigma$ is defined uniquely
by the values $p(\sigma)$, ${\rm beg}(\sigma)$ and ${\rm end}(\sigma)$, we will compute
actually for any repeat $\hat\sigma_{j}$ from~$\Psi$ the values $p(\hat\sigma_{j})$, 
${\rm beg}(\hat\sigma_{j})$ and ${\rm end}(\hat\sigma_{j})$.

Let $p$ be the minimal period of $r'$ and $r''$, and $x''$ be the prefix cyclic root of~$r''$.
Denote by~$f'$ the starting position of the leftmost cyclic root of~$r'$ which is equal to $x''$
and is not a prefix of~$r'$. Taking into account Proposition~\ref{difcyceqv}, it can be checked that
$$
f' = \begin{cases} 
{\rm beg}(r') + a(r') - a(r'') \mbox{ if } a(r')>a(r'');\\ 
{\rm beg}(r') + a(r') - a(r'') + p \mbox{ if } a(r')\le a(r''). \end{cases}
$$
Note that $f'$ has to be actually the starting position of the repeat $\hat\sigma_{k'}$
from~$\Psi$, and in this case $c(\hat\sigma_{k'})={\rm end}(r')-f'+1$. Thus, if ${\rm end}(r')-f'+1<3p$,
we can conclude that $k'=0$, otherwise $k'>0$. 

Let ${\rm end}(r')-f'+1\ge 3p$, i.e. $k'>0$.
Denote by $f_1$ the starting position of $\hat\sigma_{1}$. Note that $f_1$ is the starting position
of the rightmost cyclic root of~$r'$ which is equal to $x''$ and such that 
$c(\hat\sigma_{1})={\rm end}(r')-f_1+1\ge 3p$. Thus, taking into account Proposition~\ref{cyceqv},
we have that $f_1$ is the greatest number such that $f_1-f'$ is divisible by~$p$ and
$f_1\le {\rm end}(r')-3p+1$. Note that $f_1$ can be computed in constant time. Then
${\rm beg}(\hat\sigma_{1})=f_1$, $p(\hat\sigma_{1})={\rm beg}(r'')-f_1$ and 
${\rm end}(\hat\sigma_{1})={\rm end}(r')+p(\hat\sigma_{1})$. Note also that
$$
f'={\rm beg}(\hat\sigma_{k'})={\rm beg}(\hat\sigma_{1})-(k'-1)p=f_1-(k'-1)p,
$$
so
\begin{equation}
k'=\frac{f_1-f'}{p}+1, 
\label{eqvk'}
\end{equation}
and for any $j=1, 2,\dots , k'$ we have 
${\rm beg}(\hat\sigma_{j})={\rm beg}(\hat\sigma_{1})-p(j-1)$, 
${\rm end}(\hat\sigma_{j})={\rm end}(\hat\sigma_{1})+p(j-1)$,
$p(\hat\sigma_{j})=p(\hat\sigma_{1})+p(j-1)$ and $c(\hat\sigma_{j})=c(\hat\sigma_{1})+p(j-1)$. 
Denote $\hat k = k' + k''$. As shown above, $\hat k$ is the greatest number such that 
${\rm end}(\hat\sigma_{1})+p(\hat k-1)<{\rm end}(r'')$, so
$$
\hat k=\lfloor\frac{1}{p}({\rm end}(r'')-{\rm end}(\hat\sigma_{1})-1)\rfloor +1,
$$
and $k''=\hat k - k'$. If $k''>0$, from above observations for $j=1, 2,\ldots , k''$
we have ${\rm beg}(\hat\sigma_{k'+j})={\rm beg}(r')$, 
${\rm end}(\hat\sigma_{k'+j})={\rm end}(\hat\sigma_{k'})+jp$,
$p(\hat\sigma_{k'+j})=p(\hat\sigma_{k'})+jp$ and $c(\hat\sigma_{k'+j})=|r'|$.
Moreover, the repeat $\hat\sigma_{\hat k+1}$ from~$\Psi$ has to safisfy the conditions
${\rm beg}(\hat\sigma_{\hat k+1})={\rm beg}(r')$, ${\rm end}(\hat\sigma_{\hat k+1})={\rm end}(r'')$,
$p(\hat\sigma_{\hat k+1})=p(\hat\sigma_{\hat k})+p$. Note that in this case
$$
c(\hat\sigma_{\hat k'+1})=|\hat\sigma_{\hat k'+1}|-p(\hat\sigma_{\hat k'+1})=
{\rm end}(r'')-{\rm beg}(r')+1-p(\hat\sigma_{\hat k})-p.
$$
Thus, if
$$
{\rm end}(r'')-{\rm beg}(r')+1-p(\hat\sigma_{\hat k})-p<3p,
$$
we conclude that $k'''=0$, otherwise $k'''>0$. Let $k'''>0$. Note that
$k'''$ is the greatest number such that
$$
c(\hat\sigma_{\hat k+k'''})=c(\hat\sigma_{\hat k'+1})-p(k'''-1)\ge 3p,
$$
i.e.
$$
{\rm end}(r'')-{\rm beg}(r')+1-p(\hat\sigma_{\hat k})-pk'''\ge 3p.
$$
Thus, $k'''=\lfloor\frac{1}{p}\left({\rm end}(r'')-{\rm beg}(r')+1-p(\hat\sigma_{\hat k})\right)\rfloor -3$,
and for any $j=1, 2,\ldots , k'''$ we have ${\rm beg}(\hat\sigma_{\hat k+j})={\rm beg}(r')$,
${\rm end}(\hat\sigma_{\hat k+j})={\rm end}(r'')$, $p(\hat\sigma_{\hat k+j})=p(\hat\sigma_{\hat k+1})+(j-1)p$
and $c(\hat\sigma_{\hat k+j})=c(\hat\sigma_{\hat k+1})-(j-1)p$.

Now consider the case $k'=0$, i.e. ${\rm end}(r')-f'+1<3p$ and the pair $r', r''$ represents no repeats 
of first type. Since all repeats of second or third type represented by the pair $r', r''$ must have 
starting position ${\rm beg}(r')$, the equality ${\rm beg}(\hat\sigma_1)={\rm beg}(r')$
must be hold for $\hat\sigma_1$. Denote by $\hat\sigma_0$ the repeat with the starting position~$f'$
and the period ${\rm beg}(r'')-f'$. We proved above that $p(\hat\sigma_{k'+1})=p(\hat\sigma_{k'})+p$.
By the same way one can prove that $p(\hat\sigma_1)$ must be equal to $p(\hat\sigma_0)+p={\rm beg}(r'')-f'+p$.
In this case we obtain that $c(\hat\sigma_1)=\min (|r'|, |r''|+f'-{\rm beg}(r')-p)$.
Therefore, if $\min (|r'|, |r''|+f'-{\rm beg}(r')-p)<3p$, we conclude that $\Psi$ is empty.
Let $\min (|r'|, |r''|+f'-{\rm beg}(r')-p)\ge 3p$. Note that $f'\le {\rm beg}(r')+p$, so
${\rm end}(r')-f'+1\ge |r'|-p$. Thus, $|r'|-p<3p$. Note also that $c(\hat \sigma_{k''+2})$
has to be equal to $c(\hat \sigma_{k''+1})-p\le |r'|-p$, so $c(\hat \sigma_{k''+2})$ has
to be less than $3p$. Thus, we have $k'''\le 1$. First consider the case $|r'|<|r''|+f'-{\rm beg}(r')-p$, 
i.e. $\hat\sigma_1$ is a repeat of second type, so ${\rm end}(\hat\sigma_1)={\rm end}(r')+p(\hat\sigma_1)$.
Then $k''$ can be computed in constant time as the greatest number such that 
${\rm end}(\hat\sigma_1)+(k''-1)p<{\rm end}(r'')$, and for any $j=1, 2,\ldots , k''$ we have 
${\rm beg}(\hat\sigma_{j})={\rm beg}(r')$, ${\rm end}(\hat\sigma_{j})={\rm end}(\hat\sigma_1)+(j-1)p$,
$p(\hat\sigma_{j})=p(\hat\sigma_1)+(j-1)p$ and $c(\hat\sigma_{j})=|r'|$. Moreover, 
we obtain that $\hat\sigma_{k''+1}$ must satisfy the conditions ${\rm beg}(\hat\sigma_{k''+1})={\rm beg}(r')$, 
${\rm end}(\hat\sigma_{k''+1})={\rm end}(r'')$ and $p(\hat\sigma_{k''+1})=p(\hat\sigma_{k''})+p$, so 
$$
c(\hat\sigma_{k''+1})=|\hat\sigma_{k''+1}|-p(\hat\sigma_{k''+1})={\rm end}(r'')-{\rm beg}(r')+1
-p(\hat\sigma_{k''})-p.
$$
Thus, if ${\rm end}(r'')-{\rm beg}(r')+1-p(\hat\sigma_{k''})-p<3p$, we conclude that $k'''=0$.
Otherwise, taking into account $k'''\le 1$, we obtain that $k'''=1$ and
${\rm beg}(\hat\sigma_{k''+1})={\rm beg}(r')$, ${\rm end}(\hat\sigma_{k''+1})={\rm end}(r'')$
and $p(\hat\sigma_{k''+1})=p(\hat\sigma_{k''})+p$. Finally consider the case 
$|r'|\ge |r''|+f'-{\rm beg}(r')-p$, i.e. $k''=0$ and $\hat\sigma_1$ is a repeat of third type,
so ${\rm end}(\hat\sigma_1)={\rm end}(r'')$. Taking into account $k'''\le 1$, we obtain in this 
case that $\hat\sigma_1$ is an unique repeat in~$\Psi$. Thus, in any case $\Psi$ can be computed
in constant time.

Now we need to select from~$\Psi$ all $\alpha$-gapped repeats, i.e. all gapped repeats $\hat\sigma$
such that $\frac{p(\hat\sigma)}{c(\hat\sigma)}\le\alpha$. If $\hat k=0$, i.e. $k'=k''=0$, as shown
above, we have a trivial case when $\Psi$ contains no more than one repeat. So farther we assume
that $\hat k>0$. First we consider the case when $\hat\sigma_1$ is a gapped repeat. 
Note that for any repeats $\hat\sigma_j$ and $\hat\sigma_{j+1}$ from~$\Psi$
the ending position of the left copy of~$\hat\sigma_{j+1}$ is not greater than the ending position 
of the left copy of~$\hat\sigma_{j}$ and the starting position of the right copy of~$\hat\sigma_{j+1}$ 
is not less than the right copy of~$\hat\sigma_{j}$. Thus, if $\hat\sigma_{j}$ is a gapped repeat then
all repeats $\hat\sigma_{l}$ for $l>j$ are also gapped repeats. Therefore, in this case all repeats
from~$\Psi$ are gapped repeats. Farther we note that for any~$j$ such that $k'<j<k'+k''+k'''$ we have
$c(\hat\sigma_{j})\ge c(\hat\sigma_{j+1})$ and $p(\hat\sigma_{j})<p(\hat\sigma_{j+1})$, so
\begin{equation}
\frac{p(\hat\sigma_{k'+1})}{c(\hat\sigma_{k'+1})}<\frac{p(\hat\sigma_{k'+2})}{c(\hat\sigma_{k'+2})}<\ldots
<\frac{p(\hat\sigma_{k'+k''+k'''})}{c(\hat\sigma_{k'+k''+k'''})}.
\label{forjgrk'}
\end{equation}
Now consider a repeat $\hat\sigma_{j}$ from~$\Psi$ such that $1\le j<k'$. Since $\hat\sigma_{j}$
is a gapped repeat, we have $c(\hat\sigma_{j})<p(\hat\sigma_{j})$ and, as shown above, 
$c(\hat\sigma_{j+1})=c(\hat\sigma_{j})+p$, $p(\hat\sigma_{j+1})=p(\hat\sigma_{j})+p$.
Hence
$$
\frac{p(\hat\sigma_{j+1})}{c(\hat\sigma_{j+1})}=\frac{p(\hat\sigma_{j})+p}{c(\hat\sigma_{j})+p}<
\frac{p(\hat\sigma_{j})}{c(\hat\sigma_{j})}.
$$
Thus, we have
\begin{equation}
\frac{p(\hat\sigma_{1})}{c(\hat\sigma_{1})}>\frac{p(\hat\sigma_{2})}{c(\hat\sigma_{2})}>\dots
>\frac{p(\hat\sigma_{k'})}{c(\hat\sigma_{k'})}.
\label{forjltk'}
\end{equation}
From inequalities (\ref{forjgrk'}) and~(\ref{forjltk'}) we conclude that all the $\alpha$-gapped repeats
from~$\Psi$ have to form a continuous segment $\hat\sigma_{l}, \hat\sigma_{l+1}, \dots , \hat\sigma_{m}$
in~$\Psi$. Thus, for efficient finding of all these repeats we need to compute the indexes~$l$ and~$m$.

First we compute~$l$. Let $k'>0$ and $\frac{p(\hat\sigma_{1})}{c(\hat\sigma_{1})}\le \alpha$. Then we 
obviously obtain $l=1$. 

Now let $k'>0$ and $\frac{p(\hat\sigma_{1})}{c(\hat\sigma_{1})}>\alpha\ge 
\frac{p(\hat\sigma_{k'})}{c(\hat\sigma_{k'})}$. Then, taking into account~(\ref{forjltk'}), we have
that $l$ is smallest number such that $\frac{p(\hat\sigma_{l})}{c(\hat\sigma_{l})}\le \alpha$.
Since this number can not be greater than~$k'$, as shown above, we have also 
$p(\hat\sigma_{l})=p(\hat\sigma_{1})+p(l-1)$ and $c(\hat\sigma_{l})=c(\hat\sigma_{1})+p(l-1)$. 
Thus, $l$ is smallest number such that $\frac{p(\hat\sigma_{1})+p(l-1)}{c(\hat\sigma_{1})+p(l-1)}\le \alpha$. 
It can be easily computed that
$$
l=1+\left\lceil\frac{p(\hat\sigma_{1})-\alpha c(\hat\sigma_{1})}{p(\alpha - 1)}\right\rceil .
$$

Now let $k'=0$ or $\alpha < \frac{p(\hat\sigma_{k'})}{c(\hat\sigma_{k'})}$. In this case,
if there exists the repeat $\hat\sigma_{k'+1}$ in~$\Psi$ and 
$\frac{p(\hat\sigma_{k'+1})}{c(\hat\sigma_{k'+1})}\le\alpha$ then we have $l=k'+1$,
otherwise from~(\ref{forjgrk'}) we obtain that there are no $\alpha$-gapped repeats in~$\Psi$.

Now we compute~$m$. If $\frac{p(\hat\sigma_{\hat k+k'''})}{c(\hat\sigma_{\hat k+k'''})}\le\alpha$
then we obviously have $m=\hat k+k'''$. So farther we assume that 
$\frac{p(\hat\sigma_{\hat k+k'''})}{c(\hat\sigma_{\hat k+k'''})}>\alpha$.
Let $k'''>0$ and $\frac{p(\hat\sigma_{\hat k+1})}{c(\hat\sigma_{\hat k+1})}\le\alpha <
\frac{p(\hat\sigma_{\hat k+k'''})}{c(\hat\sigma_{\hat k+k'''})}$. Then, taking into 
account~(\ref{forjgrk'}), we have that $m$ is equal to the greatest number $\hat k + m'$
for $k'''>m'\ge 1$ such that $\frac{p(\hat\sigma_{\hat k + m'})}{c(\hat\sigma_{\hat k + m'})}\le\alpha$.
As shown above, we have also $p(\hat\sigma_{\hat k+m'})=p(\hat\sigma_{\hat k+1})+(m'-1)p$
and $c(\hat\sigma_{\hat k+m'})=c(\hat\sigma_{\hat k+1})-(m'-1)p$. Thus, $m$ is equal to 
the greatest number $\hat k + m'$ for $k'''>m'\ge 1$ such that
$\frac{p(\hat\sigma_{\hat k+1})+(m'-1)p}{c(\hat\sigma_{\hat k+1})-(m'-1)p}\le\alpha$.
It can be easily computed that
$$
m=\hat k + 1 + \left\lfloor\frac{\alpha c(\hat\sigma_{\hat k+1}) - p(\hat\sigma_{\hat k+1})}
{p(\alpha + 1)}\right\rfloor .
$$

Now let $k'''>0$ and $\frac{p(\hat\sigma_{\hat k})}{c(\hat\sigma_{\hat k})}\le\alpha <
\frac{p(\hat\sigma_{\hat k+1})}{c(\hat\sigma_{\hat k+1})}$. Then we have obviously that
$m=\hat k$. 

Now we can assume that $\alpha <\frac{p(\hat\sigma_{\hat k})}{c(\hat\sigma_{\hat k})}$
and, if $k'''>0$, $\alpha <\frac{p(\hat\sigma_{\hat k +1})}{c(\hat\sigma_{\hat k +1})}$.
Let $k''>0$ and $\frac{p(\hat\sigma_{k'+1})}{c(\hat\sigma_{k'+1})}\le\alpha <
\frac{p(\hat\sigma_{\hat k})}{c(\hat\sigma_{\hat k})}$. Then, taking into 
account~(\ref{forjgrk'}), we have that $m$ is equal to the greatest number $k' + m''$
for $k''>m''\ge 1$ such that $\frac{p(\hat\sigma_{k' + m''})}{c(\hat\sigma_{k' + m''})}\le \alpha$.
From above observations we obtain that $p(\hat\sigma_{k' + m''})=p(\hat\sigma_{k' + 1})+(m''-1)p$
and $c(\hat\sigma_{k' + m''})=|r'|$. Thus, $m$ is equal to the greatest number $k' + m''$ 
for $k''>m''\ge 1$ such that $\frac{p(\hat\sigma_{k' + 1})+(m''-1)p}{|r'|}\le\alpha$.
It can be easily computed that
$$
m=k' + 1 + \left\lfloor\frac{\alpha |r'| - p(\hat\sigma_{k'+1})}{p}\right\rfloor .
$$

Now we assume additionally that, if $k''>0$, $\alpha < \frac{p(\hat\sigma_{k'+1})}{c(\hat\sigma_{k'+1})}$
In this case, if $k'>0$ and $\alpha\ge \frac{p(\hat\sigma_{k'})}{c(\hat\sigma_{k'})}$ then $m=k'$,
othervise we can conclude that $\Psi$ has no $\alpha$-gapped repeats. Thus, $l$ and~$m$ can be
computed in constant time, so all $\alpha$-gapped repeats in~$\Psi$ can be computed in the required time.

Finally, we consider the case when $\hat\sigma_1$ is an overlapped repeat. Denote for any 
$j=1, 2,\ldots , k'+\hat k$ by $\hat u'_j$ and $\hat u''_j$ the left and right copies of
the repeat $\hat\sigma_j$. Let $k'>0$, i.e. $\hat\sigma_1$ is a repeat of first type. Note
that for any repeat $\hat\sigma_j$ of first type we have ${\rm end}(\hat u'_j)={\rm end}(\hat u'_1)$
and ${\rm beg}(\hat u''_j)={\rm beg}(\hat u''_1)$. Therefore, if $\hat\sigma_1$ is an overlapped repeat,
then any repeat $\hat\sigma_j$ of first type is also an overlapped repeat. Thus, in the considered case
we obtain that all repeats of first type from~$\Psi$ are overlapped repeats, i.e. only repeats of second 
or third types from~$\Psi$ can be gapped repeats. Now consider in~$\Psi$ a repeat $\hat\sigma_j$ such
that $j\ge k'+2$. As shown above, we have 
$$
{\rm beg}(\hat u''_j)\ge {\rm beg}(\hat u''_{k'+2})={\rm beg}(\hat u''_{k'+1})+p\ge {\rm beg}(r'')+p.
$$
On the other hand, we have ${\rm end}(\hat u'_j)\le {\rm end}(r')$. By Proposition~\ref{repoverlap0},
the overlap of repetitions $r'$ and $r''$ is less than~$p$, so ${\rm end}(r')<{\rm beg}(r'')+(p-1)$.
Thus, we obtain that ${\rm beg}(\hat u''_j)\ge {\rm beg}(r'')+p$ and ${\rm end}(\hat u'_j)<{\rm beg}(r'')+(p-1)$,
so $\hat\sigma_j$ is a gapped repeats. Hence all repeats $\hat\sigma_j$ from~$\Psi$ such that $j\ge k'+2$
are gapped repeats. Thus, all gapped repeats from~$\Psi$ are repeats $\hat\sigma_l, \hat\sigma_{l+1},\ldots ,
\hat\sigma_{k'+\hat k}$ where $l=k'+1$ if $\hat\sigma_{k'+1}$ is a gapped repeat and $l=k'+2$ otherwise.
Since one can check in constant time if the repeat $\hat\sigma_{k'+1}$ is gapped, the value~$l$ can be
computed in constant time. Taking into account inequalities~(\ref{forjgrk'}), we conclude that 
$$
\frac{p(\hat\sigma_{l})}{c(\hat\sigma_{l})}<\frac{p(\hat\sigma_{l+1})}{c(\hat\sigma_{l+1})}<\ldots
<\frac{p(\hat\sigma_{k'+\hat k})}{c(\hat\sigma_{k'+\hat k})}.
$$
Thus, if $\frac{p(\hat\sigma_{l})}{c(\hat\sigma_{l})}>\alpha$ then there are no $\alpha$-gapped repeats in~$\Psi$, 
otherwise all $\alpha$-gapped repeats in~$\Psi$ form a continuous segment $\hat\sigma_{l}, \hat\sigma_{l+1}, \dots , 
\hat\sigma_{m}$ where the index~$m$ can be computed in constant time analogously the considered above case when
$\hat\sigma_1$ is a gapped repeat. Therefore, all $\alpha$-gapped repeats in~$\Psi$ can be computed in the required 
time in any case. \qed

\begin{lemma}
Let $r', r''$ be a left pair of maximal repetitions with the same cyclic roots such that
$p(r')=p(r'')$, ${\rm beg}(r')$, ${\rm beg}(r'')$, ${\rm end}(r')$, ${\rm end}(r'')$, 
$a(r')$, $a(r'')$ be known. Then all maximal $\alpha$-gapped nondominating repeats of third type 
represented by the pair $r', r''$ can be computed in time $O(1+s)$ where $s$ is the number of the 
maximal $\alpha$-gapped nondominating repeats of third type represented by the pair $r', r''$.
\label{repnonleft}
\end{lemma}

{\bf Proof.} According to the proof of the previous lemma, all maximal $\alpha$-gapped periodic 
repeats represented by the pair $r', r''$ form a continuous segment $\hat\sigma_{l}, \hat\sigma_{l+1}, \dots , 
\hat\sigma_{m}$ in~$\Psi$ where the indexes $l, m$ can be computed in constant time. Note that
a repeat $\hat\sigma_{j}$ from~$\Psi$ is a nondominating repeat of third type if and only if 
$j>k'+k''+1$. Thus, in order to find all maximal $\alpha$-gapped nondominating repeats of third 
type represented by the pair $r', r''$, one needs to select from repeats $\hat\sigma_{l}, 
\hat\sigma_{l+1}, \dots , \hat\sigma_{m}$ all repeats $\hat\sigma_{j}$ such that $j>k'+k''+1$.
It can be done obviously in time $O(1+s)$ where $s$ is the number of the selected repeats. \qed

Let $r', r''$ be a left pair of maximal repetitions representing some $\alpha$-gapped repeat~$\sigma$
with left and right copies $u'$ and $u''$ and gap~$v$. Note that the distance 
${\rm beg}(r'')-({\rm end}(r')+1)$ between the repetitions $r'$ and $r''$ can not be
greater than the gap length $|v|$ which is not greater that $(\alpha - 1)|u'|\le (\alpha - 1)|r'|$.
Thus, ${\rm beg}(r'')-({\rm end}(r')+1)\le (\alpha - 1)|r'|$, so
$$
{\rm beg}(r'')-{\rm beg}(r')=({\rm beg}(r'')-({\rm end}(r')+1))+|r'|\le (\alpha - 1)|r'| + |r'|=\alpha |r'|.
$$
If ${\rm beg}(r'')-{\rm beg}(r')\le \alpha |r'|$, we will say that the repetition~$r''$ is 
$\alpha$-close to the repetition~$r'$ from the right. Thus, we obtain the following fact.
\begin{proposition}
If a left pair $r', r''$ of maximal repetitions represents at least one maximal $\alpha$-gapped repeat
then $r''$ is $\alpha$-close to~$r'$ from the right.
\label{alphclose}
\end{proposition}

We use also the following proposition.
\begin{proposition}
Any maximal repetition $r'$ has no more than $2\alpha$ maximal repetitions which are $\alpha$-close 
to~$r'$ from the right.
\label{nomorealpha}
\end{proposition}
{\bf Proof.} Let $r''_1, r''_2,\ldots , r''_s$ be all maximal repetitions which are $\alpha$-close 
to~$r'$ from the right in the increasing order of their starting positions:
$$
{\rm beg}(r')<{\rm beg}(r''_1)<{\rm beg}(r''_2)<\ldots < {\rm beg}(r''_s).
$$
For convenience denote $r'$ by $r''_0$ and consider two consecutive repeats $r''_j$ and $r''_{j+1}$
for $j=0, 1,\ldots , s-1$. Since $p(r''_j)=p(r''_{j+1})=p(r')$, by Proposition~\ref{repoverlap0}, the 
overlap of $r''_{j}$ and $r''_{j+1}$ is less than $p(r')$, so 
$$
{\rm beg}(r''_{j+1})>{\rm beg}(r''_{j}) + (|r''_{j}|-p(r'))\ge {\rm beg}(r''_{j}) + (|r'|-p(r'))\ge
{\rm beg}(r''_{j}) + |r'|/2.
$$
Thus, we obtain that
$$
{\rm beg}(r''_0)+s|r'|/2 < {\rm beg}(r''_{s})\le {\rm beg}(r')+\alpha |r'|={\rm beg}(r''_0)+\alpha |r'|,
$$
so $s|r'|/2 < \alpha |r'|$, i.e. $s < 2\alpha$. \qed

Farther by computing of a periodic repeat we will mean that the minimal period of copies of this repeat
is additionally computed.

\begin{lemma}
All left birepresented maximal $\alpha$-gapped periodic repeats in~$w$ can be computed in time $O(n\alpha)$.
\label{leftbirepr}
\end{lemma}

{\bf Proof.} First we find all maximal repetitions in~$w$. According to Theorem~\ref{numtimerun}, it can be done in 
$O(n)$ time, and the total number of these repetitions is $O(n)$.
For each maximal repetition~$r$ in~$w$ we compute the values $p(r)$, ${\rm beg}(r)$, ${\rm end}(r)$, $a(r)$.
Moreover, we divide all maximal repetitions in~$w$ into the subsets of all repetitions with the same cyclic roots,
and represent each of these subsets as a double-linked list $PRSR_j$ of the subset repetitions in the order of
increasing of their starting positions. According to~\cite{Crochemore14}, it can be done in $O(n)$ time. We also rearrange 
repetitions in each list $PRSR_j$ in the order of nondecreasing of their lengths (the repetitions of the
same length are arranged in the order of increasing of their starting positions). The rearranged list $PRSR_j$
will be denoted by $LRSR_j$. Using the bucket sorting, all the lists $LRSR_j$ can be computed from the lists $PRSR_j$
in total $O(n)$ time.

For each~$j$ we compute separately all maximal $\alpha$-gapped periodic repeats represented by left pairs
of repetitions from $PRSR_j$. The computation is performed as follows. We consider consecutively all repetitions
from  $LRSR_j$. For each repetition~$r'$ from  $LRSR_j$ we compute all maximal $\alpha$-gapped periodic repeats 
represented by left pairs $r', r''$ of repetitions where $r''$ is a repetition from $PRSR_j$. Before the computation 
we assume that all repetitions which precede the repetition $r'$ in the list $LRSR_j$ are already removed from 
the current list $PRSR_j$. Note that, in order to compute the required repeats for the repetition~$r'$, we need 
to consider all the repetitions~$r''$ from $PRSR_j$ such that the pair of maximal repetitions $r', r''$ is left 
and represents at least one maximal $\alpha$-gapped periodic repeat. According to Proposition~\ref{alphclose},
such repetitions~$r''$ have to be $\alpha$-close to~$r'$ from the right. Thus, we actually need to consider
only repetitions~$r''$ from $PRSR_j$ which are $\alpha$-close to~$r'$ from the right. Note that, since the
lengths of all these repetitions are not less than the length of~$r'$ and the starting positions of all these
repetitions are greater than the starting position of~$r'$, all these repetitions follows to~$r'$ in the 
list $LRSR_j$, so they are presented in the current list $PRSR_j$. Moreover, they follow to~$r'$ in $PRSR_j$.
Recall that the current list $PRSR_j$ contains no repetitions of length less than the length of~$r'$.
So in the current list $PRSR_j$ between the repetition~$r'$ and the repetitions which are $\alpha$-close to~$r'$ 
from the right there are no any other repetitions. Thus, all the repetitions which are $\alpha$-close to~$r'$ 
from the right form in the list $PRSR_j=(r''_1, r''_2,\ldots , r''_q)$ some continuous segment
$(r''_l, r''_{l+1},\ldots , r''_m)$ which follows immediately to~$r'$, i.e. $r'=r_{l-1}$.
Therefore, proceeding from the repetition~$r'$, each of the repetitions which are $\alpha$-close to~$r'$ from 
the right can be found in the current list $PRSR_j$ in constant time. After finding of each repetition~$r''$ 
from these repetitions we compute all maximal $\alpha$-gapped periodic repeats represented by the left 
pair $r', r''$. According to Lemma~\ref{repreleft}, it can be done in time $O(1+s)$ where $s$ is the number
of computed repeats. The minimal periods of copies of all these computed repeats are defined as the minimal
period of~$r'$. Thus, the treating of each of the repetitions which are $\alpha$-close to~$r'$ from 
the right can be done in time $O(1+s)$. Since, according to Lemma~\ref{repnonleft}, the number of maximal 
repetitions which are $\alpha$-close to~$r'$ from the right is not greater than $2\alpha$, the total treating
of all these repetitions can be done in time $O(\alpha +s')$ where $s'$ is the total number of computed repeats
for all these repetitions. Then we remove the considered repetition~$r'$ from the double-linked list $PRSR_j$.
It can be done in constant time. Thus, the considering of the repetition~$r'$ can be performed in time
$O(\alpha +s')$. Therefore, since the number of all maximal repetitions is $O(n)$, the total time of considering
of all maximal repetitions from all lists $LRSR_j$ is $O(n\alpha +s'')$ where $s''$ is the total number 
of computed repeats. Since each computed repeat is a maximal $\alpha$-gapped repeat, the number~$s''$ is
not greater than the total number of maximal $\alpha$-gapped repeats in~$w$ which is $O(n\alpha)$ 
by Theorem~\ref{numtimereps}, so $s''=O(n\alpha)$. Thus, the total time of considering of all maximal repetitions 
from all lists $LRSR_j$ (which is actually the time of computing all left birepresented maximal $\alpha$-gapped 
periodic repeats in~$w$) is $s''=O(n\alpha)$. \qed

Analogously to proving of Lemma~\ref{leftbirepr}, from Lemma~\ref{repnonleft} we can also prove
the following lemma.
\begin{lemma}
All left birepresented maximal $\alpha$-gapped nondominating repeats of third type in~$w$ can be computed in 
time $O(n\alpha)$.
\label{leftnonrepr}
\end{lemma}

Now consider the case of periodic repeats represented by right pairs of maximal repetitions.
Let the maximal repeat~$\sigma$ be represented by the right pair of maximal repetitions $r', r''$
with the same cyclic roots and the minimal period~$p$. Analogously to the case of repeats
represented by left pairs of repetitions, we can show that $\sigma$ can satisfy one of the
three following cases:\\
1. ${\rm end}(u')={\rm end}(r')$, ${\rm beg}(u'')={\rm beg}(r'')$ and ${\rm end}(u'')<{\rm end}(r'')$;\\
2. ${\rm beg}(u')>{\rm beg}(r')$, ${\rm beg}(u'')={\rm beg}(r'')$ and ${\rm end}(u'')={\rm end}(r'')$;\\
3. ${\rm beg}(u')={\rm beg}(r')$ and ${\rm end}(u'')={\rm end}(r'')$.\\
We will call the repeat~$\sigma$ {\it repeat of first type} in the case~1, 
{\it repeat of second type} in the case~2, and {\it repeat of third type} in the case~3.

Analogously to the case of repeats represented by left pairs of repetitions, we can also show that
starting positions of the right copies of all maximal repeats of third type represented by the pair $r', r''$
form an arithmetic progression $t''', t'''+p, t'''+2p,\ldots , t'''+(k'''-1)p$ where $k'''$ is
the number of all maximal repeats of third type represented by the pair $r', r''$. We will call
the repeat of third type with the starting position~$t'''$ of its right copy {\it dominating} repeat,
all the other repeats of third type will be called {\it nondominating}. Analogously to Proposition~\ref{prnondleft},
we can prove that a maximal repeat~$\sigma$ represented by a right pair of maximal repetitions is covered
by another repeat represented by the same pair of repetitions if and only if $\sigma$ is a nondominating 
repeat of third type. Taking it into account together with Proposition~\ref{prnondleft}, we obtain the 
following fact.
\begin{corollary}
A maximal periodic repeat~$\sigma$ represented by a pair of maximal repetitions is covered by another 
periodic repeat represented by the same pair of repetitions if and only if $\sigma$ is a nondominating 
repeat of third type.
\label{cornondom}
\end{corollary}

Analogously to Lemmas \ref{leftbirepr} and~\ref{leftnonrepr}, we can also prove the similar facts
for repeats represented by right pairs of maximal repetitions.
\begin{lemma}
All right birepresented maximal $\alpha$-gapped periodic repeats in~$w$ can be computed in time $O(n\alpha)$.
\label{rghtbirepr}
\end{lemma}
\begin{lemma}
All right birepresented maximal $\alpha$-gapped nondominating repeats of third type in~$w$ can be computed in 
time $O(n\alpha)$.
\label{rghtnonrepr}
\end{lemma}

From Lemmas \ref{leftnonrepr} and~\ref{rghtnonrepr} we obtain the following corollary.
\begin{corollary}
All birepresented maximal $\alpha$-gapped nondominating repeats of third type in~$w$ can be computed in 
time $O(n\alpha)$.
\label{nonrepr}
\end{corollary}

\begin{lemma}
All maximal $\alpha$-gapped periodic repeats in~$w$ can be computed in time $O(n\alpha)$.
\label{perepr}
\end{lemma}
{\bf Proof.} By Lemmas \ref{leftbirepr} and~\ref{rghtbirepr} we can find in time $O(n\alpha)$
all birepresented maximal $\alpha$-gapped periodic repeats in~$w$, so for finding of all maximal 
$\alpha$-gapped periodic repeats in~$w$ we need to compute additionally in~$w$ all maximal 
$\alpha$-gapped periodic repeats represented by maximal repetitions. By Proposition~\ref{repmaxrep}
maximal $\alpha$-gapped periodic repeats represented by maximal repetitions are generated by these 
maximal repetitions. So for finding of all these repeats we can compute initially all maximal 
$\alpha$-gapped repeats generated by maximal repetitions in~$w$. By Proposition~\ref{findgenreps},
the number of such repeats is $O(n)$, and all these repeats can be computed in $O(n)$ time. Note
that a maximal repeat~$\sigma$ generated by a maximal repetition~$r$ is a periodic repeat represented
by~$r$ if and only if $p(r)\le c(\sigma)/3$. Moreover, in this case $p(r)$ is the minimal period of
copies of~$\sigma$. So for each of the computed repeats we can check in constant time if this repeat 
is a periodic repeat represented by a maximal repetition and can define in this case the minimal 
period of copies of this repeat. Thus, all maximal $\alpha$-gapped repeats 
generated by maximal repetitions in~$w$ can be computed in $O(n)$ time, so the total time of 
computing of all maximal $\alpha$-gapped repeats generated by maximal repetitions in~$w$ 
is $O(n\alpha)$. \qed

%%%%%%%%%%%%%%%%%%%%%%%%%%%%%%%%%%%%%%%%%%%%%%%%%%%%%%%%%%%%%%%%%%%%%%%%%%%%%%%%%%%%%%%%%%%%%%%
\section{$\alpha$-periodic repeats}
%%%%%%%%%%%%%%%%%%%%%%%%%%%%%%%%%%%%%%%%%%%%%%%%%%%%%%%%%%%%%%%%%%%%%%%%%%%%%%%%%%%%%%%%%%%%%%%

Repeat~$\sigma$ is called {\it $\alpha$-periodic} if the minimal period of copies of~$\sigma$ is
not greater than $\frac{p(\sigma )}{3\alpha}$, otherwise $\sigma$ is called {\it $\alpha$-nonperiodic}.
Note that for any $\alpha$-periodic maximal repeat~$\sigma$ which is $\alpha$-gapped or overlapped we
have $c(\sigma)\ge p(\sigma)/\alpha\ge 3p$ where $p$ is the minimal period of copies of~$\sigma$.
So any $\alpha$-periodic maximal $\alpha$-gapped or overlapped repeat is a periodic repeat.
Thus, all $\alpha$-periodic maximal $\alpha$-gapped or overlapped repeats can be classified
as repeats represented by single maximal repetitions or repeats represented by pairs of maximal 
repetitions. 

\begin{proposition}
Let $r$ be a maximal repetition such that $e(r)\ge 3$. Then the principal repeat of~$r$
is $\alpha$-nonperiodic.
\label{princisnon}
\end{proposition}

{\bf Proof.} Let $\sigma$ be the principal repeat of~$r$. Assume that $\sigma$ is
$\alpha$-periodic, i.e. the  copies of~$\sigma$ are repetitions with minimal period~$p'$ 
not greater than $\frac{p(\sigma)}{3\alpha}=\frac{p(r)}{3\alpha}$, so $p'<p(r)/3<p(r)$.
Since $e(r)\ge 3$, i.e. $|r|\ge 3p(r)$, we have that the length $c(\sigma)$ of copies
of~$\sigma$ is not less than $2p(r)>p(r)+p'$. Therefore, since the both $p(r)$ and $p'$
are periods of copies of~$\sigma$, by periodicity lemma we obtain that $\gcd (p(r), p')$
is also a period of copies of~$\sigma$. Since $p'$ is the minimal period of copies of~$\sigma$,
we conclude that $p'=\gcd (p(r), p')$, i.e. $p'$ is a divisor of $p(r)$. In this case we
obtain that $p'$ is also a period of~$r$ which contradicts that $p(r)$ is the minimal
period of~$r$. \qed

Let $\sigma$ be a $\alpha$-periodic maximal repeat represented by some maximal repetition~$r$.
Note that in this case $\sigma$ is covered by~$r$, i.e. is covered by the principal repeat of~$r$.
Since the repeat~$\sigma$ is a periodic repeat with the minimal period $p(r)$ of its copies,
i.e. the length of its copies is not less that $3p(r)$, and $r$ contains copies of~$\sigma$,
we have that $e(r)\ge 3$. Thus, we have the following corollary from Proposition~\ref{princisnon}.
\begin{corollary}
Any $\alpha$-periodic maximal repeat represented by some maximal repetition is covered by
the $\alpha$-nonperiodic principal repeat of this repetition.
\label{periscover}
\end{corollary}

Our algorithm is based on the following lemma.

\begin{lemma}
Let a maximal $\alpha$-gapped repeat~$\sigma$ is covered by some $\alpha$-periodic maximal repeat
and is not covered by any $\alpha$-nonperiodic principal repeat of maximal repetition. Then $\sigma$
is a periodic birepresented nondominating repeat of third type.
\label{lemalphaper}
\end{lemma}

{\bf Proof.} Let $\sigma$ be covered by an $\alpha$-periodic maximal repeat~$\sigma'$.
By Proposition~\ref{onlycover}, $\sigma'$ is an $\alpha$-gapped or overlapped repeat,
so $\sigma'$ is a periodic repeat. If $\sigma'$ is represented by a maximal repetition,
then, by Corollary~\ref{periscover}, $\sigma'$ is covered by the $\alpha$-nonperiodic 
principal repeat of this repetition, so $\sigma$ is also covered by the $\alpha$-nonperiodic 
principal repeat of this repetition which contradicts conditions of the lemma.
Thus, $\sigma'$ is represented by a pair of maximal repetitions $r', r''$. Let
$p'$ the minimal period of these  repetitions. Note that, by Proposition~\ref{copycover},
the left copy of~$\sigma$ is contained in the left copy of~$\sigma'$ and
the right copy of~$\sigma$ is contained in the right copy of~$\sigma'$.
Hence the left copy of~$\sigma$ is contained in~$r'$ and the right copy of~$\sigma$ 
is contained in~$r''$, so $p'$ is a period of copies of~$\sigma$. Moreover, we have
that
$$
c(\sigma)\ge\frac{p(\sigma)}{\alpha}>\frac{p(\sigma')}{\alpha}\ge 3p'.
$$
Thus, $\sigma$ is a periodic repeat. Denote by~$p$ the minimal period of copies
of~$\sigma$. Let $p<p'$. Since $c(\sigma)>3p'>p'+p$, by periodicity lemma
we have that $\gcd (p', p)$ is also a period of copies of~$\sigma$ which has
to be equal to~$p$. Thus, $p$ is a divisor of $p'$, so in the case $p<p'$
we obtain that the cyclic roots of repetitions $r', r''$ are not primitive.
Therefore, $p'=p$, i.e. $p'$ is the minimal period of copies of~$\sigma$, so
$\sigma$ is represented by the pair of repetitions $r', r''$. Thus, we have
that $\sigma$ is a periodic repeat represented by the pair $r', r''$ and is
covered by the periodic repeat $\sigma'$ represented by the same pair $r', r''$.
So, by Corollary~\ref{cornondom}, $\sigma$ is a nondominating repeat of third type. \qed

\begin{lemma}
All maximal $\alpha$-gapped $\alpha$-periodic repeats in~$w$ can be computed 
in time $O(n\alpha)$.
\label{bialprepr}
\end{lemma}
{\bf Proof.} Recall that any maximal $\alpha$-gapped $\alpha$-periodic repeat is a periodic 
repeat, so for finding of all maximal $\alpha$-gapped $\alpha$-periodic repeats we can
compute initially all maximal $\alpha$-gapped periodic repeats in~$w$. By Lemma~\ref{perepr},
it can be done in time $O(n\alpha)$. Moreover, by Theorem~\ref{numtimereps}, the number of
computed repeats is $O(n\alpha)$. Recall that for each computed repeat we compute additionally
the minimal period of copies of this repeat, so we can check in constant time if this repeat
is $\alpha$-periodic. Thus, in $O(n\alpha)$ time we can select from the computed repeats all
maximal $\alpha$-gapped $\alpha$-periodic repeats in~$w$. The total time of the proposed
procedure for computing of the required repeats is $O(n\alpha)$. \qed

%\begin{corollary}
%All maximal $\alpha$-gapped $\alpha$-nonperiodic repeats in~$w$ can be computed 
%in time $O(n\alpha)$.
%\label{binonalprepr}
%\end{corollary}
%{\bf Proof.} First we find all maximal $\alpha$-gapped repeats in~$w$. By Theorem~\ref{findgapreps},
%the number of these repeats is $O(n\alpha)$ and all these repeats can be found in $O(n\alpha)$ time.
%Further we find all maximal $\alpha$-gapped $\alpha$-periodic repeats in~$w$. By Lemma~\ref{bialprepr},
%it can be done also in $O(n\alpha)$ time. Then, in order to find all maximal $\alpha$-gapped 
%$\alpha$-nonperiodic repeats in~$w$, we remove from all maximal $\alpha$-gapped repeats all
%maximal $\alpha$-gapped $\alpha$-periodic repeats in~$w$. It can be done in the following way.

Now we consider maximal overlapped periodic repeats represented by pairs of maximal repetitions.
Note that such repeats can be represented only by pairs of overlapped maximal repetitions.
\begin{lemma}
Let $r', r''$ be a left pair of maximal overlapped repetitions with the same cyclic roots 
such that $p(r')=p(r'')$, ${\rm beg}(r')$, ${\rm beg}(r'')$, ${\rm end}(r')$, ${\rm end}(r'')$, 
$a(r')$, $a(r'')$ be known. Then the number of maximal overlapped periodic repeats represented 
by the pair $r', r''$ is less than $e(r')$, and all these repeats can be computed in time $O(e(r'))$.
\label{reoverleft}
\end{lemma}
{\bf Proof.} It is shown in the proof of Lemma~\ref{repreleft} that in the set~$\Psi$ all repeats 
$\hat\sigma_j$ such that $j\ge k'+2$ are gapped repeats, so only repeats $\hat\sigma_1, \hat\sigma_2,
\ldots , \hat\sigma_{k'+1}$ can be overlapped repeats, and, moreover, all these $k'+1$ repeats can be
computed in $O(1+k')$ time. It is also shown that repeats $\hat\sigma_1, \hat\sigma_2,\ldots , \hat\sigma_{k'}$
are overlapped if and only if repeat $\hat\sigma_1$ is overlapped. Thus, in constant time we can select
from repeats $\hat\sigma_1, \hat\sigma_2,\ldots , \hat\sigma_{k'+1}$ all overlapped repeats.
It follows from equation~(\ref{eqvk'}) that $k'\le\frac{|r'|}{p(r')}-2=e(r')-2$. Thus, the
number the selected repeats is not greater than $1+k'<e(r')$, and the total time of computing
of these repeats is $O(1+k')=O(e(r'))$. \qed

\begin{lemma}
The number of maximal overlapped periodic repeats represented by left pairs of maximal repetitions
is $O(n)$, and all these repeats can be  computed in $O(n)$ time.
\label{leftoverepr}
\end{lemma}
{\bf Proof.} Analogously to the proof of Lemma~\ref{leftbirepr}, first we find all maximal repetitions in~$w$, 
for each maximal repetition~$r$ in~$w$ we compute the values $p(r)$, ${\rm beg}(r)$, ${\rm end}(r)$, $a(r)$, and,
moreover, we divide all maximal repetitions in~$w$ into the subsets of all repetitions with the same cyclic roots
and represent each of these subsets as a list $PRSR_j$ of the subset repetitions in the order of increasing of 
their starting positions. As shown in the proof of Lemma~\ref{leftbirepr}, it can be done in $O(n)$ time.
Then we consider each list $PRSR_j$. Let $PRSR_j$ consists of consecutive repetitions 
$(r_1^{(j)}, r_2^{(j)},\ldots , r_q^{(j)})$. According to Proposition~\ref{repoverlap0}, for each 
repetition $r_i^(j)$ the overlaps of $r_i^{(j)}$ with repetitions $r_{i-1}^{(j)}$ and $r_{i+1}^{(j)}$ 
are less than $p(r_i^{(j)})$, so repetitions $r_{i-1}^{(j)}$ and $r_{i+1}^{(j)}$
can not be overlapped. Thus, only pairs $r_i^{(j)}, r_{i+1}^{(j)}$ can be overlapped pairs of repetitions 
representing maximal overlapped repeats. Therefore, we traverse the list $PRSR_j$ for finding all overlapped 
left pairs $r_i^{(j)}, r_{i+1}^{(j)}$ of repetitions. Note that the total number of maximal repetitions in~$w$
is $O(n)$, so the total time of traversing of all lists $PRSR_j$ is $O(n)$. For each found pair $r_i^{(j)}, r_{i+1}^{(j)}$ 
we compute all maximal overlapped periodic repeats represented by this pair. By Lemma~\ref{reoverleft}, 
the number of these repeats is less than $e(r_i^{(j)})$, and all these repeats can be computed in time $O(e(r_i^{(j)}))$.
For the computed repeats the minimal periods of copies of these repeats are defined as the minimal
period of~$r_i^{(j)}$. Note that for each maximal repetition~$r$ in~$w$ there can be only one left pair 
$r_i^(j), r_{i+1}^{(j)}$ of repetitions such that $r\equiv r_i^{(j)}$, so the total number of maximal 
overlapped periodic repeats represented by all left pairs $r_i^{(j)}, r_{i+1}^{(j)}$ of repetitions in all 
lists $PRSR_j$ is less than $\sum_{r\in {\cal R}(w)} e(r)$, so, by Theorem~\ref{sumexp}, this number is $O(n)$. 
By the same reason, the total time of computing of all maximal overlapped periodic repeats represented 
by all left pairs $r_i^{(j)}, r_{i+1}^{(j)}$ of repetitions in all lists $PRSR_j$ is $O(\sum_{r\in {\cal R}(w)} e(r))$,
so, by Theorem~\ref{sumexp}, this time is $O(n)$. Thus, the total time of the considered procedure 
for computing of the required repeats is $O(n)$. \qed

We can also prove the analogous lemma for right pairs of maximal repetitions.
\begin{lemma}
The number of maximal overlapped periodic repeats represented by right pairs of maximal repetitions
is $O(n)$, and all these repeats can be  computed in $O(n)$ time.
\label{rghtoverepr}
\end{lemma}

From Lemmas \ref{leftoverepr} and~\ref{rghtoverepr} we directly obtain the corollary.
\begin{corollary}
The number of maximal overlapped periodic birepresented repeats is $O(n)$, and all these repeats can be  
computed in $O(n)$ time.
\label{coroverepr}
\end{corollary}

\begin{lemma}
The number of maximal overlapped $\alpha$-periodic birepresented repeats is $O(n)$, and all these repeats 
can be computed in $O(n)$ time.
\label{lemalphrepr}
\end{lemma}
{\bf Proof.} Recall that any maximal overlapped $\alpha$-periodic repeat is a periodic repeat, 
so for finding of all maximal overlapped $\alpha$-periodic birepresented repeats we can
compute initially all maximal overlapped periodic birepresented repeats in~$w$. By Corollary~\ref{coroverepr},
the number of such repeats is $O(n)$, and all these repeats can be computed in $O(n)$ time.
Since for each computed repeat we compute additionally the minimal period of copies of this repeat, 
we can check in constant time if this repeat is $\alpha$-periodic. Thus, in $O(n)$ time we can select 
from the computed repeats all maximal overlapped $\alpha$-periodic birepresented repeats in~$w$. 
The total time of the proposed procedure for computing of the required repeats is $O(n)$. \qed

\begin{lemma}
The number of reprincipal $\alpha$-periodic birepresented repeats is $O(n)$, and all these repeats 
can be computed in $O(n)$ time.
\label{prinalphrepr}
\end{lemma}
{\bf Proof.} Recall that all reprincipal repeats are maximal overlapped repeats, so for computing of
all reprincipal $\alpha$-periodic birepresented repeats in~$w$ we can select them from all maximal 
overlapped $\alpha$-periodic birepresented repeats in~$w$. By Lemma~\ref{lemalphrepr}, the number 
of all maximal overlapped $\alpha$-periodic birepresented repeats is $O(n)$, and these repeats 
can be computed in $O(n)$ time. By Proposition~\ref{findprincip} the number of reprincipal repeats in~$w$
is also $O(n)$, and all these repeats can be computed in $O(n)$ time. In order to select all
reprincipal repeats from maximal overlapped $\alpha$-periodic birepresented repeats, we represent 
the set all the computed maximal overlapped $\alpha$-periodic birepresented repeats by start position
lists $MOBRL_t$. By the same way, we represent the set of all the computed reprincipal repeats by 
start position lists $PRL_t$. All the lists $MOBRL_t$ and $PRL_t$ can be computed in time $O(n+S)$
where $S$ is the total size of all lists $MOBRL_t$ and $PRL_t$, so, since this total size is $O(n)$,
the time of computing of all the lists $MOBRL_t$ and $PRL_t$ is $O(n)$. Then, in order to select
the required reprincipal repeats, we traverse simultaneously lists $MOBRL_t$ and $PRL_t$ for 
$t=1, 2,\ldots , n$. It can be also done in time $O(n+S)=O(n)$. Thus, the total time of the
proposed procedure for computing of the required repeats is $O(n)$. \qed

Now we consider reprincipal repeats represented by maximal repetitions.
\begin{proposition}
The principal repeat of a maximal repetition can not be represented by 
another maximal repetition.
\label{princirep}
\end{proposition}
{\bf Proof.} Let $\sigma$ be the principal periodic repeat of a maximal repetition~$r$.
Assume that $\sigma$ is represented by some another maximal repetition~$r'$. Note that
in this case $p(r')\le p(r)$ and $r$ is contained in $r'$, i.e. the length of the overlap
of the repetitions $r$ and $r'$ is not less than $2p(r)$. It contradicts Proposition~\ref{repoverlap1}. \qed

\begin{corollary}
Any reprincipal $\alpha$-periodic repeat is a birepresented repeat.
\label{corprinalph}
\end{corollary}
{\bf Proof.} Let $\sigma$ be the principal $\alpha$-periodic repeat of some maximal repetition~$r$.
Note that, as shown above, $\sigma$ is a periodic repeat. Assume that $\sigma$ is represented by
some maximal repetition. By Proposition~\ref{princirep}, $\sigma$ can be represented only by repetition~$r$.
In this case we have that $p(\sigma)=p(r)$ and the minimal period of copies of~$\sigma$ is $p(r)$,
so the minimal period of copies of~$\sigma$ is greater than $\frac{p(\sigma )}{3\alpha}$ which
contradicts that $\sigma$ is a $\alpha$-periodic repeat. \qed

From Lemma~\ref{prinalphrepr} and Corollary~\ref{corprinalph} we obtain immediately the following fact.
\begin{corollary}
The number of reprincipal $\alpha$-periodic repeats is $O(n)$, and all these repeats can be computed in 
$O(n)$ time.
\label{coralphprin}
\end{corollary}

%%%%%%%%%%%%%%%%%%%%%%%%%%%%%%%%%%%%%%%%%%%%%%%%%%%%%%%%%%%%%%%%%%%%%%%%%%%%%%%%%%%%%%%%%
\section{Algorithm for solving of Problem~\ref{thirdprob}}
%%%%%%%%%%%%%%%%%%%%%%%%%%%%%%%%%%%%%%%%%%%%%%%%%%%%%%%%%%%%%%%%%%%%%%%%%%%%%%%%%%%%%%%%%

Let $\alpha = 1/\delta$. We compute initially the set $GR$ of all maximal $\alpha$-gapped 
repeats in~$w$ and the set $PR$ of all reprincipal repeats in~$w$. By Theorem~\ref{numtimereps},
we have that $|GR|=O(n\alpha)$ and $GR$ can be computed in $O(\alpha n)$ time.
Moreover, by Proposition~\ref{findprincip}, we have that $|PR|=O(n)$ and $PR$ can be 
computed in $O(n)$ time. Recall that our goal is to exclude from $GR$ all repeats 
which are covered by other maximal repeats. Using Propositions \ref{onlycover} 
and~\ref{overcover}, we conclude that we need actually to exclude from $GR$ all 
repeats which are covered by other maximal gapped repeats from $GR$ or reprincipal 
repeats. Denote by $GR^*$ the set of all repeats from $GR$ which are not covered 
by other repeats from $GR$ or reprincipal repeats. In these terms, our goal is 
to compute the set $GR^*$.

Note that all maximal repeats generated by a maximal repetition except the
principal repeat of this repetition are covered by this repetition and so
are covered by the principal repeat of this repetition. Recall also that
principal repeats of maximal repetitions are overlapped repeats, so gapped
repeats cannot be reprincipal repeats. Thus, maximal gapped repeats generated 
by a maximal repetition are covered by the principal repeat of this repetition 
and so have to be excluded from $GR$. At the first stage we exclude from $GR$ 
all maximal $\alpha$-gapped repeats generated by maximal repetitions. 
By Proposition~\ref{findgenreps}, the number of these repeats is $O(n)$, and all 
these repeats can be computed in $O(n)$ time. Denote the set of all the computed 
repeats by $CR$.
%Note that any maximal repetition~$r$ generates no more 
%than $e(r)$ maximal gapped repeats, and these repeats can be easily computed 
%from~$r$ for $O(e(r))$ time. Thus, taking into account Theorem~ref{sumexp},
%we have that the number of maximal gapped repeats generated by maximal 
%repetitions in~$w$ is $O(n)$, and all these repeats can be computed in $O(n)$
%time from the already computed maximal repetitions in~$w$. 
In order to exclude from $GR$ the repeats of the set $CR$, we represent $GR$ by 
start position lists $GRL_t$. These lists can be computed in time $O(n+|GR|)=O(\alpha n)$. 
By the same way, we represent the set $CR$ by start position lists $CRL_t$. 
These lists can be computed in time $O(n+|CR|)=O(n)$. Then all the computed repeats 
which is contained in $GR$ can be excluded from $GR$ by simultaneous traversing of 
lists $GRL_t$ and $CRL_t$ in time $O(n+|GR|+|CR|)=O(\alpha n)$.

Denote by $GR'$ the resulting set of all repeats from $GR$ which remain
after the first stage. Recall that all repeats from $GR$ which are removed
at the first stage are covered by reprincipal repeats. Therefore, any repeat
from $GR$ which is covered by a repeat~$\sigma$ removed at the first stage
is covered also by some reprincipal repeat covering the repeat~$\sigma$.
Thus, in order to compute the set $GR^*$, we can remove from $GR'$ all
repeats which are covered by other repeats from $GR'$ or reprincipal repeats.
Denote by $\widehat {GR}$ the set of all repeats from $GR'$ together with
all reprincipal repeats in~$w$. In these terms, in order to compute the set $GR^*$,
we remove from $GR'$ all repeats which are covered by other repeats from $\widehat {GR}$.

At the second stage we remove from $GR'$ all repeats which are covered
by $\alpha$-nonperiodic repeats from $\widehat {GR}$. For this purpose,
first we compute the set $AGR$ of all $\alpha$-periodic repeats from $GR$
and the set $APR$ of all $\alpha$-periodic reprincipal repeats in~$w$.
By Lemma~\ref{bialprepr}, the set $AGR$ can be computed in $O(n\alpha)$ time,
and, by Corollary~\ref{coralphprin}, the set $APR$ can be computed in $O(n)$ 
time. Note that after performing of the first stage the set $GR'$ is represented
by its start position lists $GRL'_t$. The set $AGR$ can be also represented by
start position lists $AGRL_t$ which can be computed in $O(n+|AGR|)=O(n\alpha)$ time.
Then, using simultaneous traversing of lists $GRL'_t$ and $AGRL_t$, we mark all
$\alpha$-nonperiodic repeats from $GR'$ as $\alpha$-nonperiodic (each repeat 
from $GRL'_t$ which is not contained in $AGRL_t$ is marked as $\alpha$-nonperiodic).
Moreover, we compute the set $NPR$ of all $\alpha$-nonperiodic reprincipal repeats 
in~$w$ by removing from the set $PR$ all repeats from $APR$. To perform thus removing,
we also represent the sets $PR$ and $APR$ by their start position lists $PRL_t$ 
and $APRL_t$. All these lists can be computed in time $O(n+|PR|+|APR|)=O(n)$.
Then we compute all repeats from $NPR$ by simultaneous traversing of lists $PRL_t$ 
and $APRL_t$ in total time $O(n+|PR|+|APR|)=O(n)$. Note that the computed set $NPR$
is also represented by its start position lists $NPRL_t$. Then we compute the set
$\widehat {GR'}=GR'\cup NPR$ by merging the lists $GRL'_t$ and $NPRL_t$ into the
start position lists $\widehat {GRL'}_t$ for this set. It can be done by simultaneous 
traversing of lists $GRL'_t$ and $NPRL_t$ in time $O(n+|GR'|+|NPR|)=O(\alpha n)$.
During the merging of repeats into the lists $\widehat {GRL'}_t$ we also mark these
repeats as gapped or as reprincipal. Note also that $|\widehat {GR'}|=|GR'|+|NPR|=O(\alpha n)$.

%Note that these repeats are not generated by maximal
%repetitions. Thus, in order to find all these repeats, we can consider
%the set $\widehat {GR'}$ of all repeats from $GR$ which are not generated 
%by maximal repetitions together with all principal $\alpha$-nonperiodic 
%repeats of maximal repetitions in~$w$. In this set we need to find all
%gapped repeats covered by other $\alpha$-nonperiodic repeats from this
%set. We suppose that all repeats from $\widehat {GR'}$ are sorted by their 
%starting positions, i.e. for each starting position $t=1, 2,\ldots, n$ 
%all repeats from $\widehat {GR'}$ with the starting position~$t$ are 
%contained in the list $GRL_t$ in the order of increasing of their periods.
%We also suppose that in the considered lists $GRL_t$ all $\alpha$-nonperiodic 
%gapped repeats and all principal repeats of maximal repetitions are specially
%marked. Farther for convenience we will also call principal repeats of 
%maximal repetitions as principal repeats.

For $i=1,2,\ldots , \lfloor\log_2(n-1)\rfloor$ we denote by $RQ_i$ the
subset of all repeats~$\sigma$ from $\widehat {GR'}$ such that $2^i\le p(\sigma)<2^{i+1}$,
i.e. $\lfloor\log_2 p(\sigma)\rfloor = i$.

\begin{proposition} Let a maximal gapped repeat~$\sigma$ be covered
by a  maximal gapped repeat~$\sigma'$. Then $p(\sigma)\ge p(\sigma')> p(\sigma)/2$.
\end{proposition}

{\bf Proof.} Note that $p(\sigma)<|\sigma |\le |\sigma'|<2p(\sigma')$, so $p(\sigma')> p(\sigma)/2$.
The inequality $p(\sigma)\ge p(\sigma')$ is obvious. \qed

\begin{corollary} Let a gapped repeat~$\sigma$ from $RQ_i$ be covered by a gapped repeat~$\sigma'$
from $\widehat {GR'}$. Then $\sigma'$ is contained in $RQ_i$ or $RQ_{i-1}$.
\label{corsigmafri}
\end{corollary}

\begin{proposition} Let a gapped repeat~$\sigma$ from $RQ_i$ be covered by a reprincipal repeat~$\sigma'$ 
from $RQ_{i'}$. Then $i\ge i'\ge i-\lceil\log_2\alpha\rceil$.
\label{propsigmarqi}
\end{proposition}

{\bf Proof.} It is obvious that $i\ge i'$. Assume that $i'<i-\lceil\log_2\alpha\rceil$. Note that
for any~$\sigma$ from $RQ_i$ and any~$\sigma'$ from $RQ_{i'}$ we have $p(\sigma )>2^{i-i'-1}p(\sigma')$,
so in this case $p(\sigma )>2^{\lceil\log_2\alpha\rceil}p(\sigma')\ge\alpha p(\sigma')$.
Hence, $c(\sigma )\ge p(\sigma )/\alpha >p(\sigma')$. Let $u'$, $u''$ be left and right copies of $\sigma$.
Since $\sigma$ is covered by $\sigma'$, the repeat $\sigma$ is contained in the repetition
${\rm rep}(\sigma')$ with the minimal period $p(\sigma')$, so the both copies $u'$, $u''$ are contained 
in ${\rm rep}(\sigma')$ and $|u'|=|u''|>p(\sigma')$. Consider the prefixes of length $p(\sigma')$
in $u'$ and $u''$. Since these prefixes are equal cyclic roots of ${\rm rep}(\sigma')$, by Proposition~\ref{cyceqv}
we have ${\rm beg}(u')\equiv {\rm beg}(u'')\pmod{p(\sigma')}$, so $p(\sigma)={\rm beg}(u'')-{\rm beg}(u')$
is divisible by $p(\sigma')$, i.e. $p(\sigma)$ is divisible by the minimal period of ${\rm rep}(\sigma')$.
Moreover, if ${\rm beg}(u')>{\rm beg}({\rm rep}(\sigma'))$ then both symbols $w[{\rm beg}(u')-1]$ and
$w[{\rm beg}(u'')-1]$ are contained in ${\rm rep}(\sigma')$, so $w[{\rm beg}(u')-1]=w[{\rm beg}(u'')-1]$,
i.e. $\sigma$ can be extended to the left which contradicts that $\sigma$ is a maximal repeat. Thus, 
${\rm beg}(u')={\rm beg}({\rm rep}(\sigma'))$. It can be analogously proved that 
${\rm end}(u'')={\rm end}({\rm rep}(\sigma'))$. Thus, the repeat $\sigma$ is generated by the repetition 
${\rm rep}(\sigma')$ which contradicts that $\sigma\in\widehat {GR'}$. \qed

Summing up Corollary~\ref{corsigmafri} and Proposition~\ref{propsigmarqi}, we obtain
the following fact.
\begin{corollary}
Let a gapped repeat~$\sigma$ from $RQ_i$ be covered by a repeat~$\sigma'$ from $RQ_{i'}$. 
Then $i\ge i'\ge i-\lceil\log_2\alpha\rceil$.
\label{corsigmarqi}
\end{corollary}

For finding all repeats which have to be removed at the second stage, for each starting 
position~$t=1, 2,\ldots , n$ we compute consecutively such repeats starting at position~$t$. 
Note that such repeats starting at position~$t$ can be covered only by $\alpha$-nonperiodic 
repeats from $\widehat {GR'}$ which are starting at position not greater than~$t$ and ending 
at position greater than~$t$. Denote the set of all these $\alpha$-nonperiodic repeats by $SQ$ 
and put $SQ_i=SQ\cap RQ_i$ for $i=1,2,\ldots , \lfloor\log_2(n-1)\rfloor$. Note that if for some repeat~$\sigma$
from $SQ_i$ there exists a repeat~$\sigma'$ in $SQ$ such that ${\rm end} (\sigma)\le {\rm end} (\sigma')$
and $p(\sigma')\le p(\sigma)$ then $\sigma$ can be excluded from consideration. The remaining
repeats from $SQ_i$ form a sequence $LQ_i=\sigma_1, \sigma_2,\ldots, \sigma_s$ such that
${\rm end} (\sigma_1)<{\rm end} (\sigma_2)<\ldots <{\rm end} (\sigma_s)$ and 
$p(\sigma_1)<p(\sigma_2)<\ldots <p(\sigma_s)$. In order to perform effective search in this
sequence, we present $LQ_i$ as AVL-tree $LQT_i$. For each~$i$ we compute also the value ${\rm lep}_i$
which is the maximum of ending positions of the last repeats in sequences $LQ_{i-1}$, $LQ_{i-2}$,\ldots , 
$LQ_{i-\lceil\log_2\alpha\rceil}$.

Let $LQ_i=\sigma_1, \sigma_2,\ldots, \sigma_s$. For any repeat $\sigma$ we define 
$pe(\sigma)=p(\sigma)+{\rm end} (\sigma)$. From ${\rm end} (\sigma_1)<{\rm end} (\sigma_2)<\ldots <{\rm end} (\sigma_s)$ 
and $p(\sigma_1)<p(\sigma_2)<\ldots <p(\sigma_s)$ we have $pe(\sigma_1)<pe(\sigma_2)<\ldots <pe(\sigma_s)$.

\begin{lemma} 
For each $j\le s-2$ the inequality $pe(\sigma_{j+2})>pe(\sigma_{j})+\frac{p(\sigma_{j})}{3\alpha}$ is valid.
\label{lemjles-2}
\end{lemma}

{\bf Proof.} Denote for convenience by $u'_1$, $u'_2$, $u'_3$ the left copies of $\sigma_{j}$,
$\sigma_{j+1}$, $\sigma_{j+2}$, and by $u''_1$, $u''_2$, $u''_3$ the right copies of $\sigma_{j}$,
$\sigma_{j+1}$, $\sigma_{j+2}$. For $k=1, 2, 3$ denote $p_k=p(\sigma_{j+k-1})$, $e_k={\rm end} (\sigma_{j+k-1})$,
$pe_k=p_k+e_k$ and for $k=2, 3$ denote also $\Delta p_k=p_{k}-p_1$, $\Delta e_k=e_{k}-e_1$,
$\Delta pe_k=pe_{k}-pe_1$.

Assume that $pe_3\le pe_1 + \frac{p_1}{3\alpha}$, i.e. $\Delta pe_2 < \Delta pe_3\le\frac{p_1}{3\alpha}$.
Consider separately the following three cases.

1. Let $u''_1$ is contained in $u''_2$. So $u'_2$ contains the factor $\hat u'_1$ corresponding
to the factor $u''_1$ in $u''_2$ such that ${\rm end} (u'_1)-{\rm end} (\hat u'_1)=\Delta p_2$. Thus,
since $u'_1=\hat u'_1$, we obtain that $u'_1$ has the period $\Delta p_2<\Delta pe_2<\frac{p_1}{3\alpha}$
which contradicts that $\sigma_{j}$ is $\alpha$-nonperiodic.

2. Now let $u''_1$ is not contained in $u''_2$, i.e. ${\rm beg}(u''_1)<{\rm beg}(u''_2)$,
and $\Delta e_2\le \Delta p_2$ which implies ${\rm end}(u'_2)\le {\rm end}(u'_1)$. Let $u''$ be the
intersection of $u''_1$ and $u''_2$. Since $u''$ is a factor of $u''_1$ and $u''_2$, for $u''$ there 
are corresponding factors $u'$ in $u'_1$ and $\hat u'$ in $u'_2$. Since ${\rm end}(u')-{\rm end}(\hat u')=\Delta p_2$
and $u'=\hat u'=u''$, the both factors $u'$ and $\hat u'$ have the period $\Delta p_2$. Note that
$\Delta p_2$ and $\Delta e_2$ are less than $\Delta pe_2 < \frac{p_1}{3\alpha}$. Thus,
\begin{equation}
|u''|=|u''_2| - \Delta e_2 > |u''_2| - \frac{p_1}{3\alpha}\ge \frac{p_2}{\alpha} - \frac{p_1}{3\alpha}>
\frac{p_1}{\alpha} - \frac{p_1}{3\alpha}= \frac{2p_1}{3\alpha} > 2 \Delta p_2.
\label{relforu2}
\end{equation}
Therefore, the intersection of $u'$ and $\hat u'$ has length greater than $\Delta p_2$, so,
by Proposition~\ref{unionper}, the union $\hat u$ of factors $u'$ and $\hat u'$ has also the period $\Delta p_2$.
Note that $u'_2$ is contained in $\hat u$, so $u'_2$ has also the period $\Delta p_2<\frac{p_1}{3\alpha}<\frac{p_2}{3\alpha}$ 
which contradicts that $\sigma_{j+1}$ is $\alpha$-nonperiodic.

3. Finally, let $u''_1$ is not contained in $u''_2$ and $\Delta e_2>\Delta p_2$ which implies
 ${\rm end}(u'_2)>{\rm end}(u'_1)$. As in case~2, we define the factors $u''$ and $u'$,
and show that both these factors have period $\Delta p_2$. Note that $u''$ is a prefix of $u''_1$,
so $u'$ is a prefix of $u'_1$. Thus, ${\rm end}(u')={\rm end}(u'_1)$ . Consider the symbol 
$w[{\rm end}(u'_1)+1]$ following the copy $u'_1$. Since $\sigma_1$ is maximal, we have
$w[{\rm end}(u'_1)+1]\neq w[{\rm end}(u''_1)+1]$. Moreover, since ${\rm end}(u'_2)>{\rm end}(u'_1)$,
the symbol $w[{\rm end}(u'_1)+1]$ is contained in $u'_2$, so $u''_2$ contains the corresponding symbol
$w[{\rm end}(u'_1)+p_2+1]=w[{\rm end}(u''_1)+\Delta p_2+1]$ which is equal to $w[{\rm end}(u'_1)+1]$.
Thus, we obtain that $w[{\rm end}(u''_1)+1]\neq w[{\rm end}(u''_1)+\Delta p_2+1]$. Let $w[{\rm end}(u''_1)+1]$
be not contained in $u''_3$, i.e. ${\rm beg}(u''_3)>{\rm end}(u''_1)+1$. In this case we have that
$$
\Delta e_3>|u''_3|\ge\frac{p_3}{\alpha}>\frac{p_1}{\alpha}
$$
which contradicts that $\Delta e_3<\Delta pe_3\le\frac{p_1}{3\alpha}$. Thus, $w[{\rm end}(u''_1)+1]$ is contained 
in $u''_3$. Therefore, since the symbol $w[{\rm end}(u''_1)+\Delta p_2+1]$ is contained in $u''_2$
and ${\rm end}(u''_2)<{\rm end}(u''_3)$, the symbol $w[{\rm end}(u''_1)+\Delta p_2+1]$ is also
contained in $u''_3$. Thus, $u''_3$ contains the both unequal symbols $w[{\rm end}(u''_1)+1]$
and $w[{\rm end}(u''_1)+\Delta p_2+1]$. Therefore, $u''_3$ contains the corresponding unequal 
symbols $w[{\rm end}(u''_1)+1-p_3]$ and $w[{\rm end}(u''_1)+\Delta p_2+1-p_3]$ which can be also
represented as $w[{\rm end}(u'_1)-\Delta p_3+1]$ and $w[{\rm end}(u'_1)-\Delta p_3+\Delta p_2+1]$.
Let ${\rm end}(u'_1)-\Delta p_3+1\ge {\rm beg}(u')$. Note that $\Delta p_3>\Delta p_2$, so
${\rm end}(u'_1)-\Delta p_3+\Delta p_2+1\le {\rm end}(u'_1)$. Thus, in this case we have
$$
{\rm beg}(u')\le {\rm end}(u'_1)-\Delta p_3+1 < {\rm end}(u'_1)-\Delta p_3+\Delta p_2+1\le {\rm end}(u'_1)={\rm end}(u'),
$$
i.e. the both unequal symbols $w[{\rm end}(u'_1)-\Delta p_3+1]$ and $w[{\rm end}(u'_1)-\Delta p_3+\Delta p_2+1]$
are contained in $u'$ which contradicts that $u'$ has period $\Delta p_2$. Thus,
$$
{\rm end}(u'_1)-\Delta p_3+1<{\rm beg}(u')={\rm end}(u')-|u'|+1={\rm end}(u'_1)-|u'|+1,
$$
so $\Delta p_3>|u'|=|u''|$. By relation~(\ref{relforu2}), we have $|u''|>\frac{2p_1}{3\alpha}$.
Thus, $\Delta p_3>\frac{2p_1}{3\alpha}$ which contradicts that $\Delta p_3<\frac{p_1}{3\alpha}$.

Since we obtained contradictions in all considered cases, the lemma is proved. \qed

\begin{lemma} 
${\rm end}(\sigma_{s-1})<t+2^{i+2}$.
\label{lemsigmas-1}
\end{lemma}

{\bf Proof.} Assume by contradiction that ${\rm end}(\sigma_{s-1})\ge t+2^{i+2}$. Note that
in this case $|\sigma_{s-1}|>2^{i+2}>2p(\sigma_{s-1})$, so $\sigma_{s-1}$ is a overlapped repeat.
Since ${\rm end}(\sigma_{s})>{\rm end}(\sigma_{s-1})$, by the same reason, $\sigma_{s}$ is also 
a overlapped repeat. Note that the intersection of repetitions ${\rm rep}(\sigma_{s})$ and ${\rm rep}(\sigma_{s-1})$
contains the factor $w[t..t+2^{i+2}]$, so the overlap of these repetitions is greater than
$2^{i+2}>p(\sigma_{s-1})+p(\sigma_{s})$ which contradicts Proposition~\ref{repoverlap1}, i.e. the lemma is proved. \qed

Using Lemma~\ref{lemsigmas-1}, we have that
$$
t+2^i<pe(\sigma_1)<pe(\sigma_2)<\ldots<pe(\sigma_{s-1})<t+3\cdot 2^{i+1},
$$
and, by Lemma~\ref{lemjles-2}, for each $j<s-2$ we have 
$$
pe(\sigma_{j+2})-pe(\sigma_{j})>\frac{p(\sigma_{j})}{3\alpha}\ge\frac{2^i}{3\alpha}.
$$
Thus, we obtain that $s-1=O(\alpha )$, so $s=O(\alpha )$. Here we state this fact.

\begin{corollary}
$|LQ_i|=O(\alpha )$ for any~$i$.
\label{LQibound}
\end{corollary}

The step of the algorithm for a starting position~$t$ is as follows. First we remove from trees $LQT_i$ 
all repeats ending at position~$t$ if such repeats exist. In order to perform effectively this removing, 
for each starting position~$t$ we can maintain a double linked list containing all ending at position~$t$ 
repeats from trees $LQT_i$. In this case all repeats which have to be removed from trees $LQT_i$ can be 
found in time proportional to the number of these repeats. Then we consider consecutively all repeats in 
list $\widehat{GRL'}_t$. Let $\sigma$ be a current considered gapped repeat from $\widehat{GRL'}_t$, 
and $\lfloor\log_2 p(\sigma)\rfloor = i$. By Corollary~\ref{corsigmafri}, $\sigma$ can be covered only
by repeats from $RQ_{i'}$ where $i\ge i'\ge i-\lceil\log_2\alpha\rceil$. Note that $\sigma$ is covered 
by a repeat from $LQT_{i'}$ such that $i-1\ge i'\ge i-\lceil\log_2\alpha\rceil$ if and only if 
${\rm end} (\sigma )\le {\rm lep}_{i}$. Thus, in this case we remove $\sigma$ from $\widehat{GRL'}_t$. 
Otherwise, we check if $\sigma$ is covered by a repeat from $LQT_{i}$. For this purpose we search in $LQT_{i}$ 
the maximal~$k$ such that $p(\sigma_k)<p(\sigma )$. If such~$k$ exist and ${\rm end} (\sigma )\le 
{\rm end} (\sigma_k)$ then $\sigma$ is covered by $\sigma_k$, so in this case we also remove $\sigma$ 
from $\widehat{GRL'}_t$. Otherwise, if $\sigma$ is a $\alpha$-nonperiodic repeat, we insert $\sigma$ in $LQT_{i}$ 
between $\sigma_k$ and $\sigma_{k+1}$. After that we remove from $LQT_{i}$ all $\sigma_j$ such that 
$j>k$ and ${\rm end} (\sigma_j)\le {\rm end} (\sigma)$. If it is necessary, we update the values 
${\rm lep}_{i+1}$, ${\rm lep}_{i+2}$,\ldots , ${\rm lep}_{i+\lceil\log_2\alpha\rceil}$ which can 
depend on the value ${\rm end} (\sigma)$. Now let $\sigma$ be a current considered reprincipal repeat 
from $\widehat{GRL'}_t$. In this case we remove $\sigma$ from $\widehat{GRL'}_t$ and insert $\sigma$ in $LQT_{i}$
where $i = \lfloor\log_2 p(\sigma)\rfloor$ by the same way as we insert above a $\alpha$-nonperiodic 
gapped repeat which is checked to be not covered. Then we proceed to the next repeat in $\widehat{GRL'}_t$.

Note that during the described procedure for each repeat~$\sigma$ from $\widehat {GR'}$ we perform
at most one search in some tree $LQT_{i}$, at most one insertion of~$\sigma$ in $LQT_{i}$ and
at most one deletion of~$\sigma$ from $LQT_{i}$. All these operations can be performed in
$O(\log |LQ_i|)$ time. Since, by Corollary~\ref{LQibound}, $|LQ_i|=O(\alpha )$, we obtain
that all these operations can be performed in $O(\log \alpha)$ time. Moreover, after insertion
of~$\sigma$ no more than $\lceil\log_2\alpha\rceil$ values ${\rm lep}_{i}$ can be updated.
It can be also performed in $O(\log \alpha)$ time. All the other operations over~$\sigma$
required for the described procedure can be performed in constant time. Thus, each repeat from $\widehat {GR'}$
can be treated in $O(\log \alpha )$ time, so the total time of the described procedure for
all starting positions~$t$ is $O(n+|\widehat {GR'}|\log\alpha)$. Recall that $|\widehat {GR'}|=O(n\alpha)$, 
so the total time of the described procedure is $O(n\alpha\log\alpha)$.

Note that after the second stage we removed from $\widehat {GR'}$ all reprincipal repeats and
all repeats from $GR'$ which are covered by $\alpha$-nonperiodic repeats from $\widehat {GR}$,
so after the second stage the set $\widehat {GR'}$ consists of all repeats from $GR'$ which are not
covered by $\alpha$-nonperiodic repeats from $\widehat {GR}$. Note also that $|\widehat {GR'}|=O(n\alpha)$.

At the third stage we compute the set $GR^*$ by removing from the set $\widehat {GR'}$ all repeats
which are not contained in $GR^*$. Let $\sigma$ be a repeat from $\widehat {GR'}$ which be not
contained in $GR^*$, i.e. $\sigma$ be covered some other repeat from $\widehat {GR}$.
Since $\widehat {GR'}$ consists of repeats which are not covered by $\alpha$-nonperiodic repeats 
from $\widehat {GR}$, the repeat $\sigma$ can be covered only by an $\alpha$-periodic repeat
from $\widehat {GR}$. Moreover, $\sigma$ can not be covered by any $\alpha$-nonperiodic reprincipal  
repeat, since, otherwise, $\sigma$ has to be removed from $\widehat {GR'}$ at the second stage.
Thus, $\sigma$ is covered by some $\alpha$-periodic repeat from $\widehat {GR}$ and is not covered
by any $\alpha$-nonperiodic reprincipal repeat. Therefore, by Lemma~\ref{lemalphaper}, $\sigma$ is 
a periodic birepresented nondominating repeat of third type. On the other hand, if $\sigma$ is 
a periodic birepresented nondominating repeat of third type from $\widehat {GR'}$, then, by
Corollary~\ref{cornondom}, $\sigma$ is covered by another periodic repeat represented by the same 
pair of repetitions, so $\sigma$ is not contained in $GR^*$. Thus, a repeat from $\widehat {GR'}$
is not contained in $GR^*$ if and only if this repeat is a periodic birepresented nondominating 
repeat of third type. Hence for computing the set $GR^*$ we remove from $\widehat {GR'}$ all
periodic birepresented nondominating repeats of third type. Recall that $\widehat {GR'}$ consists
of maximal $\alpha$-gapped repeats, so we need actually to remove from $\widehat {GR'}$ all
periodic birepresented maximal $\alpha$-gapped nondominating repeat of third type. For this purpose,
we compute the set $BANR$ of all periodic birepresented maximal $\alpha$-gapped nondominating repeats 
of third type in~$w$. By Corollary~\ref{nonrepr}, this set can be computed in time $O(n\alpha)$.
Moreover, since the set $BANR$ is a subset of the set $GR$, we have $|BANR|=O(n\alpha)$. Then
we represent the set $BANR$ by its start position lists $BANRL_t$ in time $O(n+|BANR|)=O(n\alpha)$.
Finally, we remove from $\widehat {GR'}$ all repeats from $BANR$ by simultaneous traversing of lists 
$\widehat{GRL'}_t$ and $BANRL_t$ in time $O(n+|\widehat {GR'}|+|BANR|)=O(\alpha n)$. As a result, we
obtain the required set $GR^*$. The total time of the third stage procedure is $O(n\alpha)$.

Summing up the times of all the procedures of the proposed algorithm, we obtain that the total
time of the algorithm is $O(n\alpha\log\alpha)=O(\frac{n}{\delta}\log\frac{1}{\delta})$. Thus,
Problem~\ref{thirdprob} can be resolved in $O(\frac{n}{\delta}\log\frac{1}{\delta})$ time.
Since, as shown above, Problem~\ref{thirdprob} is equivalent to Problem~\ref{firstprob},
we conclude the following main result of our paper.
\begin{theorem}
All  maximal $\delta$-subrepetitions in a given word of length~$n$ over integer alphabet can be 
found in time $O(\frac{n}{\delta}\log\frac{1}{\delta})$.
\end{theorem}

%%%%%%%%%%%%%%%%%%%%%%%%%%%%%%%%%%%%%%%%%%%%%%%%%%%%%%%%%%%%%%%%%%%%%%%%%%%%%%%%%%%%%%%%%
\section{Conclusion}
%%%%%%%%%%%%%%%%%%%%%%%%%%%%%%%%%%%%%%%%%%%%%%%%%%%%%%%%%%%%%%%%%%%%%%%%%%%%%%%%%%%%%%%%%

In the paper we proposed an algorithm for finding of all  maximal $\delta$-subrepetitions 
in a given word of length~$n$ in time $O(\frac{n}{\delta}\log\frac{1}{\delta})$. 
By Proposition~\ref{numdelsubs}, the number of all maximal $\delta$-subrepetitions 
in a word of length~$n$ is $O(\frac{n}{\delta})$, so the considered problem could be presumably
resolved in $O(\frac{n}{\delta})$ time. Thus, finding of all maximal $\delta$-subrepetitions in 
a given word of length~$n$ in time $O(\frac{n}{\delta})$ is still an open problem.

\end{document}